\documentclass[prl,twocolumn,showpacs,nofootinbib,amsmath,amssymb]{revtex4-1}
\usepackage{graphicx,bm}
\usepackage{color}

\begin{document}
\title{Tomographic Constraints on High-Energy Neutrinos of Hadronuclear
Origin}
\author{Shin'ichiro Ando}
\affiliation{GRAPPA Institute, University of Amsterdam, 1098 XH
Amsterdam, The Netherlands}
\author{Irene Tamborra}
\affiliation{GRAPPA Institute, University of Amsterdam, 1098 XH
Amsterdam, The Netherlands}
\author{Fabio Zandanel}
\affiliation{GRAPPA Institute, University of Amsterdam, 1098 XH
Amsterdam, The Netherlands}
\date{September 14, 2015; revised October 22, 2015}

\begin{abstract}
 Mounting evidence suggests that the TeV--PeV neutrino flux detected by
 the IceCube telescope has mainly an extragalactic origin.
 If such neutrinos are primarily produced by a single class of
 astrophysical sources via hadronuclear ($pp$) interactions, a similar
 flux of gamma-ray photons is expected.
 For the first time, we employ {\it tomographic constraints} to pinpoint
 the origin of the IceCube neutrino events
 by analyzing recent measurements of the cross correlation between the
 distribution of GeV gamma rays, detected by the Fermi satellite, and
 several galaxy catalogs in different redshift ranges. We find that the
 corresponding bounds on the neutrino luminosity density
 are up to one order of magnitude tighter than
 those obtained by using only the spectrum of the gamma-ray background,
 especially for sources with mild redshift evolution.
 In particular, our method excludes any hadronuclear source 
 with a spectrum softer than $E^{-2.1}$
 as a main component of the neutrino background, 
 if its evolution is slower than $(1+z)^3$.
 Starburst galaxies, if able to accelerate and
 confine cosmic rays efficiently, satisfy both spectral and
 tomographic constraints.
\end{abstract}
\pacs{95.85.Pw, 95.85.Ry, 98.70.Rz, 98.70.Vc}
\maketitle

{\em Introduction}.---%
The  discovery of the PeV neutrinos by IceCube~\cite{Aartsen:2013bka,Aartsen:2013jdh} 
has launched the era of high-energy neutrino astronomy.
The current data set is compatible with a flux in excess with respect to
the atmospheric background, with an isotropic allocation of
events on the celestial sphere and flavor
equipartition~\cite{Aartsen:2013bka, Aartsen:2013jdh, Aartsen:2014gkd,
Aartsen:2014muf, Aartsen:2015ita, Aartsen:2015rwa}. 
Due to the current low statistics, the origin of the high-energy IceCube
events is not yet known, but an extragalactic and mostly diffuse
origin appears to be favored~\cite{Anchordoqui:2013dnh,Waxman:2013zda}.

The high-energy neutrino production from cosmic accelerators has been
subject of a cascade of theoretical studies, especially after the
IceCube results were announced~\cite{Anchordoqui:2013dnh,
Waxman:2013zda}.
Many papers discuss the neutrino emission from one specific source class
by adopting a model-dependent approach, for active galactic nuclei
(AGNs)~\cite{Stecker:1991vm,Stecker:2005hn, Kalashev:2013vba, Stecker:2013fxa, 
Murase:2014foa, Kimura:2014jba, Dermer:2014vaa, Kalashev:2014vya,
Khiali:2015tfa, Padovani:2015mba}, star-forming 
galaxies~\cite{Loeb:2006tw, Thompson:2006np, He:2013cqa, Liu:2013wia,
Chang:2014hua, Tamborra:2014xia,
Anchordoqui:2014yva,Chakraborty:2015sta, Senno:2015tra, Bartos:2015xpa}, gamma-ray
bursts~\cite{Waxman:1998yy,Meszaros:2001ms, Murase:2005hy, Gupta:2006jm, Murase:2006mm,
Liu:2012pf, Murase:2013ffa, Tamborra:2015qza}, galaxy clusters~\cite{Murase:2008yt, Kotera:2009ms, Murase:2012rd,
Zandanel:2014pva}, and  dark matter decays~\cite{Feldstein:2013kka,
Esmaili:2013gha,Esmaili:2014rma, Zavala:2014dla, Murase:2015gea}.

Alternatively, a more generic approach focuses on the phenomenological
aspects of the potential sources.
For example, assuming photomeson production ($p\gamma$) of neutrinos, 
Ref.~\cite{Winter:2013cla} obtained constraints on the source size and
magnetic field strength needed to match the IceCube flux.
Reference~\cite{Murase:2013rfa} hypothesized
that the TeV--PeV neutrinos were generated via hadronuclear interactions
($pp$) and concluded that the cosmic ray spectrum of the dominant
neutrino sources should be harder than $E^{-2.2}$.
This is because the associated gamma-ray spectrum will extend down to
GeV energies, where the flux of the isotropic gamma-ray background
(IGRB) measured with the Fermi Large Area
Telescope (LAT)~\cite{Ackermann:2014usa} cannot be overshot.
The connection with sources of ultrahigh-energy cosmic rays
has also been considered~\cite{Anchordoqui:2013qsi,
Fang:2014uja, Kistler:2013my}.

In this {\em Letter}, we complement the existing model-independent investigations of $pp$
neutrino sources by proposing an entirely new method: {\it Tomographic constraints}, 
up to now adopted  in studying IGRB sources.
We base this approach  on the measurements of Ref.~\cite{Xia:2015wka},
which analyzed the IGRB data and found that they were
spatially correlated with galaxy distributions.
Compared to the commonly adopted spectral analysis, the tomographic
method allows to efficiently extract a dominant IGRB component in
certain redshift ranges following galaxy catalogs, as
originally proposed for dark matter detection~\cite{Ando:2013xwa,
Ando:2014aoa, Fornengo:2013rga}. This
 provides stringent constraints on 
astrophysical sources~\cite{Xia:2015wka} and dark matter~\cite{Cuoco:2015rfa}.

We show that the tomographic approach allows to tightly constrain the redshift 
evolution and the energy spectrum of any class of astrophysical source producing
high-energy neutrinos through $pp$ interactions,
especially if the source luminosity density mildly evolves as a
function of redshifts.
It provides constraints on the expected neutrino flux that are
more stringent by up to one order of magnitude with respect to the 
common spectral approach (e.g., Ref.~\cite{Murase:2013rfa}).
We find that any source with a spectrum softer than $E^{-2.1}$ is
excluded, if its redshift evolution is slower than $(1+z)^3$.
On the other hand, sources with hard spectrum and fast
evolution can still be dominant in both gamma-rays and neutrinos.

Besides the $pp$ origin of the high-energy neutrinos, we assume
that: (i) the energy spectrum is a power law, $E^{-\alpha}$, extending
up to PeV energies; (ii) the source luminosity density evolves
as $(1+z)^\delta$ up to $z_c$, and is constant for $z > z_c$; and
(iii) the astrophysical sources trace the underlying dark matter distribution.
The third assumption is generic for any known extragalactic
source that is likely associated with large cosmic structures. 
We adopt cosmological parameters from Ref.~\cite{Ade:2015xua}.

{\em Gamma-ray intensity}.---%
We first introduce a differential gamma-ray intensity, $I_\gamma(E)$,
as the number of gamma-ray photons received per unit area, unit time,
unit solid angle, and unit energy.
It is computed as an integral of the gamma-ray window function,
$W_\gamma(E,z)$, over the comoving distance $\chi$:
\begin{eqnarray}
I_\gamma(E) &=& \int d\chi \, W_\gamma(E,z)\ , \\
W_\gamma(E,z) &=& \frac{1}{4\pi  \Lambda E_{\rm min}^{\prime 2}}
\left(\frac{E}{E'_{\rm min}}\right)^{-\alpha}
\frac{n(z)\langle L_\gamma(z)\rangle}{(1+z)^\alpha}  e^{-\tau(E,z)}\ ,
\nonumber\\
 \label{eq:window gamma}
\end{eqnarray}
where $n(z)$ is the source number density at $z$, $\langle
L_\gamma (z) \rangle$ is the mean gamma-ray luminosity emitted between
$E'_{\rm min} = 0.1$~GeV and $E'_{\rm max} = 100$~GeV in the source rest
frame (as represented by $'$), and
\begin{equation}
 \Lambda = \left\{
\begin{array}{ccc}
\frac{1-(E'_{\rm max}/E'_{\rm min})^{2-\alpha}}{\alpha-2} & \mbox{for } &
 \alpha\neq 2\ ,\\
\ln(E'_{\rm max}/E'_{\rm min}) & \mbox{for } & \alpha = 2\ .
\end{array}
\right. 
\end{equation}
The source luminosity density is assumed to evolve as
\begin{equation}
n(z)\langle L_\gamma(z)\rangle = \mathcal{E}_{\gamma,0} \times
\left\{\begin{array}{ccc}
(1+z)^\delta & \mbox{for } & z \le z_c\ , \\
(1+z_c)^\delta & \mbox{for } & z > z_c\ .
\end{array}\right.
\end{equation}
The constant evolution above $z_c$ is motivated by the observations of
infrared luminosity density of star-forming galaxies
(e.g.,~\cite{Gruppioni:2013jna}).
We note that unless the redshift dependence continues to increase
steeply up to high $z$, our conclusions are largely unaffected.
Very-high-energy gamma rays are subject to absorption by the
extragalactic background light (EBL). This is taken into account through
the exponential term in Eq.~(\ref{eq:window gamma}), where $\tau(E,z)$
is the optical depth~\cite{Finke:2009xi}.

For each set of $(\alpha, \, \delta, \, z_c)$, by taking
$\mathcal{E}_{\gamma,0}$ as a free parameter, we compute the $\chi^2$
statistic as follows:
\begin{equation}
 \chi^2 = \sum_i \left(\frac{I_{i,{\rm dat}}-I_{i,{\rm th}}(\mathcal
		  E_{\gamma, 0}|\alpha,\delta,z_c)}{\sigma_{i,{\rm
		  dat}}}\right)^2 ,
\end{equation}
where $I_{i,{\rm dat}}$ and $\sigma_{i,{\rm dat}}$ are the spectral
intensity data and the associated root-mean-square error in the $i$-th
energy bin, respectively, and $I_{i,{\rm th}}(\mathcal{E}_{\gamma,0})$
is the theoretical model intensity for $\mathcal{E}_{\gamma,0}$.
The 95\% confidence level (CL) upper limit on $\mathcal{E}_{\gamma,0}$
is obtained by solving $\Delta \chi^2 = \chi^2 - \chi_{\rm min}^2 =
2.71$.

\begin{figure}
 \begin{center}
  \includegraphics[width=8.5cm]{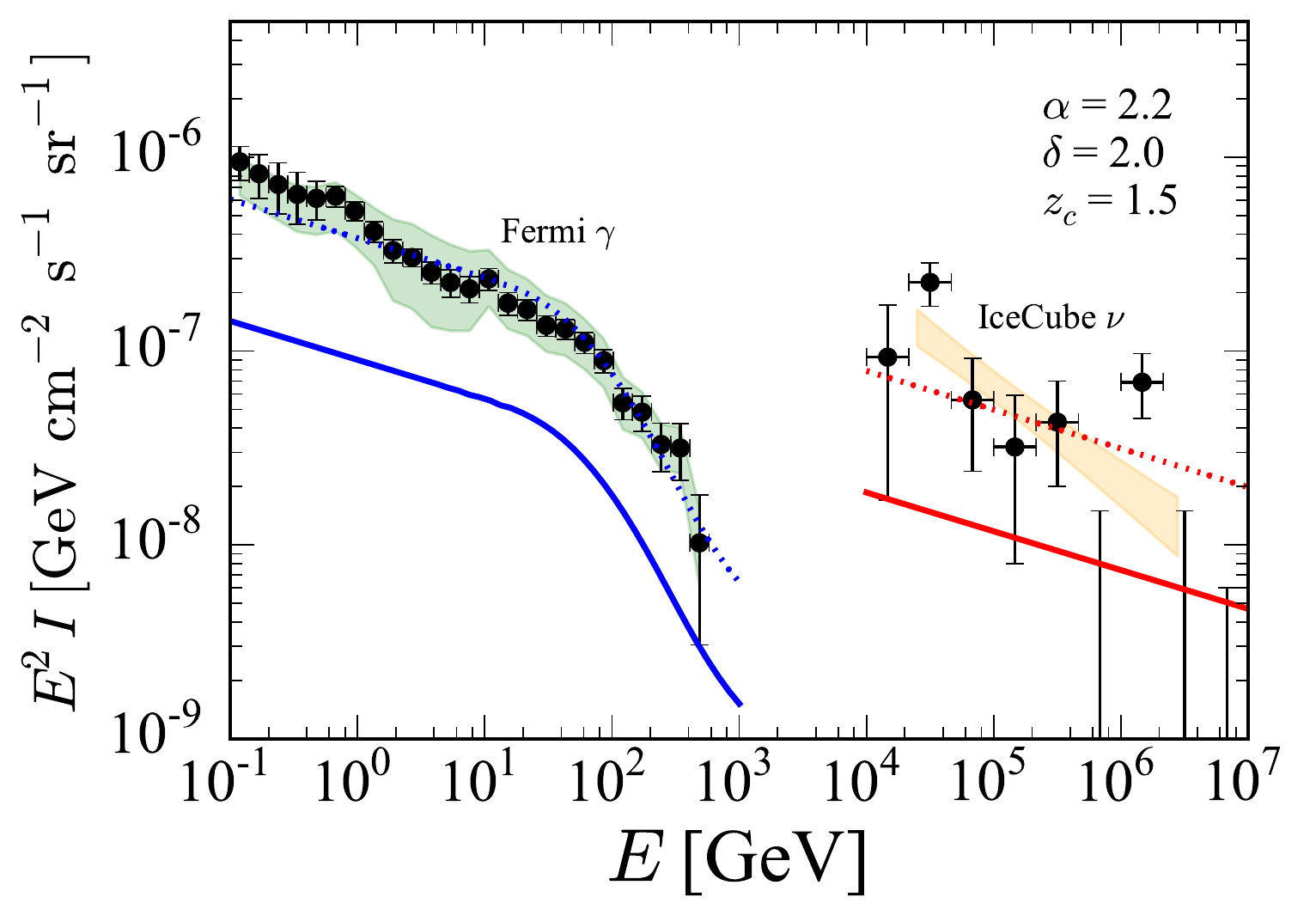}
  \includegraphics[width=8.5cm]{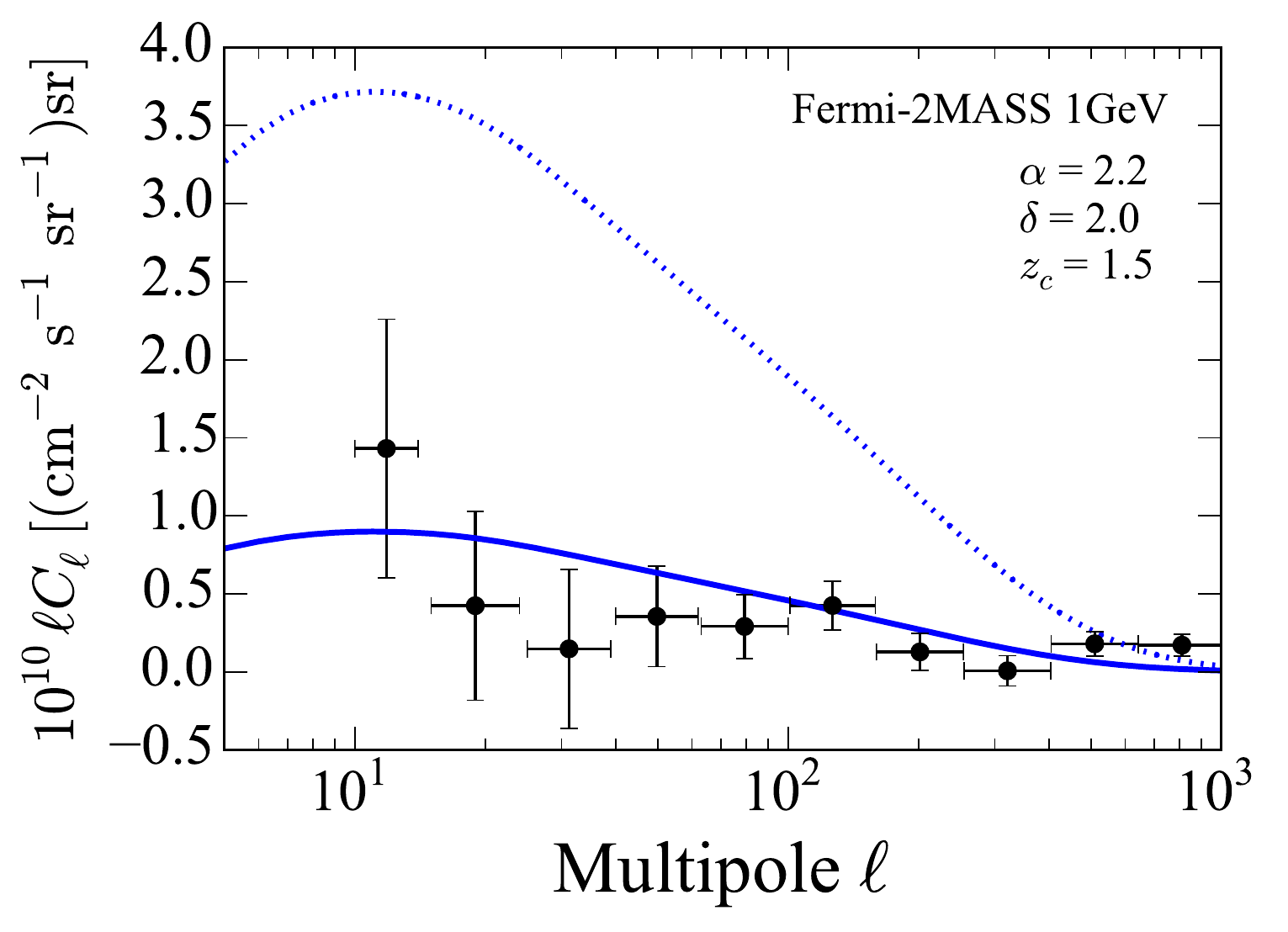}
  \caption{{\it Top:} Gamma-ray (blue) and neutrino (red) intensities
  for a model with $\alpha = 2.2$, $\delta = 2$, and $z_c = 1.5$. The dotted
  curves correspond to the 95\% CL upper limit due to the Fermi spectrum
  data (the green band represents the systematic uncertainty due to the subtraction of the
  Galactic emission~\cite{Ackermann:2014usa}). The solid curves correspond
  to the same limit but due to the cross-correlation data. IceCube data
  for the neutrino intensity are shown above 10~TeV, whereas the orange band 
  represents the 68\% CL region of the corresponding best-fit single power-law model~\cite{Aartsen:2015ita}. 
  {\it Bottom:}
  Cross-correlation angular power spectrum between the Fermi data, above
  1~GeV, and the 2MASS galaxies, compared with the
  measurements by Ref.~\cite{Xia:2015wka}. Model parameters as well as line
  types are the same as the top panel.}
  \label{fig:spectrum}
 \end{center}
\end{figure}

The top panel of Fig.~\ref{fig:spectrum} shows the gamma-ray
spectrum for $\alpha = 2.2$, $\delta = 2$, and $z_c = 1.5$ (blue
dotted), compared with the IGRB measured by
Fermi~\cite{Ackermann:2014usa}.
The value of the local luminosity density $\mathcal{E}_{\gamma,0}$
corresponding to the 95\% CL upper limit is
$\mathcal{E}_{\gamma,0}^{95\%\,\mathrm{CL}} = 2.5\times 10^{45} ~\mathrm{erg~yr^{-1}~Mpc^{-1}}$.

{\em Cross correlation with galaxy catalogs}.---%
The cross-correlation angular power spectrum, $C_\ell^{\gamma {\rm g}}$,
between the gamma-ray intensity, $I_\gamma (\hat{\bm{n}})$, and the galaxy surface density, $\Sigma_{\rm
g}(\hat{\bm n})$, is related to the angular correlation function through
the following relation (e.g.,~\cite{Ando:2014aoa}):
\begin{equation}
 \langle \delta I_\gamma(\hat{\bm n}) \, \delta \Sigma_{\rm g}(\hat{\bm n}
  + \bm\theta)\rangle = \sum_\ell\frac{2\ell+1}{4\pi}\,C_\ell^{\gamma {\rm
  g}} \, \mathcal{W}_\ell \, P_\ell(\cos\theta)\ ,
\end{equation}
where $\delta I_\gamma = I_\gamma - \langle I_\gamma \rangle$, $\delta
\Sigma_{\rm g} = \Sigma_{\rm g} - \langle \Sigma_{\rm g} \rangle$,
$P_\ell (\cos\theta)$ is the Legendre polynomial, and $\mathcal{W}_\ell$
is the beam window function (i.e., the Legendre transform of the point
spread function of the Fermi-LAT~\cite{Xia:2015wka}).

The angular cross-power spectrum $C_\ell^{\gamma {\rm
g}}$ is computed as (e.g.,~\cite{Ando:2014aoa})
\begin{equation}
 C_\ell^{\gamma {\rm g}} = \int \frac{d\chi}{\chi^2} W_\gamma (z) W_{\rm
  g}(z) P_{\gamma {\rm g}}\left(k=\frac{\ell}{\chi},z\right)\ ,
\end{equation}
where $W_{\gamma}(z)$ is the integrated gamma-ray window function, and
$W_{\rm g}(z)$ is the galaxy window function that is related to the
galaxy redshift distribution, $dN_{\rm g}/dz$, via $W_{\rm g}(z) = (d\ln N_{\rm
g} / dz)(dz / d\chi)$.
We approximate the cross-correlation power spectrum between the gamma-ray 
emitters and the galaxy catalogs as $P_{\gamma {\rm g}} \approx b_{\gamma} b_{\rm g}  P_{\rm m}$,
where $P_{\rm m}$ is the nonlinear matter power spectrum computed with the
publicly available {\tt CLASS} code~\cite{Blas:2011rf}, and $b_{\rm g}$
and $b_{\gamma}$ are the bias factors for the catalog galaxies and the
gamma-ray emitters, respectively.
We assume that gamma-ray sources are unbiased tracers of the dark matter
distribution, i.e., $b_\gamma = 1$.
Since astrophysical sources are typically positively biased dark matter
tracers (e.g.,~\cite{Xia:2015wka} and references therein),
it is a conservative assumption.

The cross-correlation analysis of Ref.~\cite{Xia:2015wka} adopted
five different catalogs: Two Micron All Sky Survey (2MASS), quasars in
the Sloan Digital Sky Survey (SDSS), the SDSS main galaxy sample, luminous
red galaxies in SDSS, and radio galaxies in the NRAO VLA Sky Survey
(NVSS).
Each of these catalogs traces underlying dark matter distribution in
a certain redshift range with a characteristic bias $b_{\rm g}$ as in
Ref.~\cite{Xia:2015wka}.
Although some of them represent AGNs, they can be used the same way as
galaxies, for which we call them ``galaxy'' catalogs collectively.
We use a redshift distribution $dN_{\rm g}/dz$ and a typical bias
$b_{\rm g}$ appropriate for each catalog, and
 three different energy ranges for the gamma rays
($>500$~MeV, $>1$~GeV, and $>10$~GeV)~\cite{Xia:2015wka}.

Similarly to the spectral analysis, for each given set of $(\alpha, \,
\delta, \, z_c)$, we compute the $\chi^2$ as follows:
\begin{equation}
 \chi^2 = \sum_{\gamma,{\rm g}} \sum_{\ell,\ell'}
  \left(C_{\rm dat}^{\gamma {\rm g}} - C_{\rm th}^{\gamma {\rm
   g}}\right)_\ell
  \left(\mbox{Cov}^{-1}\right)_{\ell\ell'}
  \left(C_{\rm dat}^{\gamma {\rm g}} - C_{\rm th}^{\gamma {\rm
   g}}\right)_{\ell'},
\end{equation}
where $\gamma$ and g run through three energy bins and five galaxy
catalogs, respectively, $\ell$ and $\ell'$ represent the multipole bins
of the measurements, and Cov is the covariance matrix.
We again use $\Delta \chi^2 = 2.71$ as a criterion to obtain the 95\% CL
upper limit on $\mathcal{E}_{\gamma,0}$.

In the bottom panel of Fig.~\ref{fig:spectrum}, we show, with a solid curve, 
the $C_\ell^{\gamma {\rm g}}$ corresponding to the 95\% CL upper limit
for $\alpha = 2.2$, $\delta = 2$, and $z_c = 1.5$,
compared with the cross-correlation data between the $>1$~GeV
photons and the 2MASS galaxies, which gives the major contribution to the
$\chi^2$. The dotted curve, in contrast, is the 95\% CL upper limit due to the
spectral data alone, and it is clearly inconsistent with the cross-correlation measurement.
The top panel of the same figure shows the corresponding energy spectra
for both approaches. It is clear that the source with the parameters adopted 
in Fig.~\ref{fig:spectrum} cannot be the main component of the IGRB
spectrum because the cross correlation provides a tighter constraint:
$\mathcal{E}_{\gamma,0}^{95\%\, \mathrm{CL}} = 5.9 \times
10^{44} ~\mathrm{erg~yr^{-1}~Mpc^{-3}}$.

\begin{figure}
 \begin{center}
  \includegraphics[width=8.5cm]{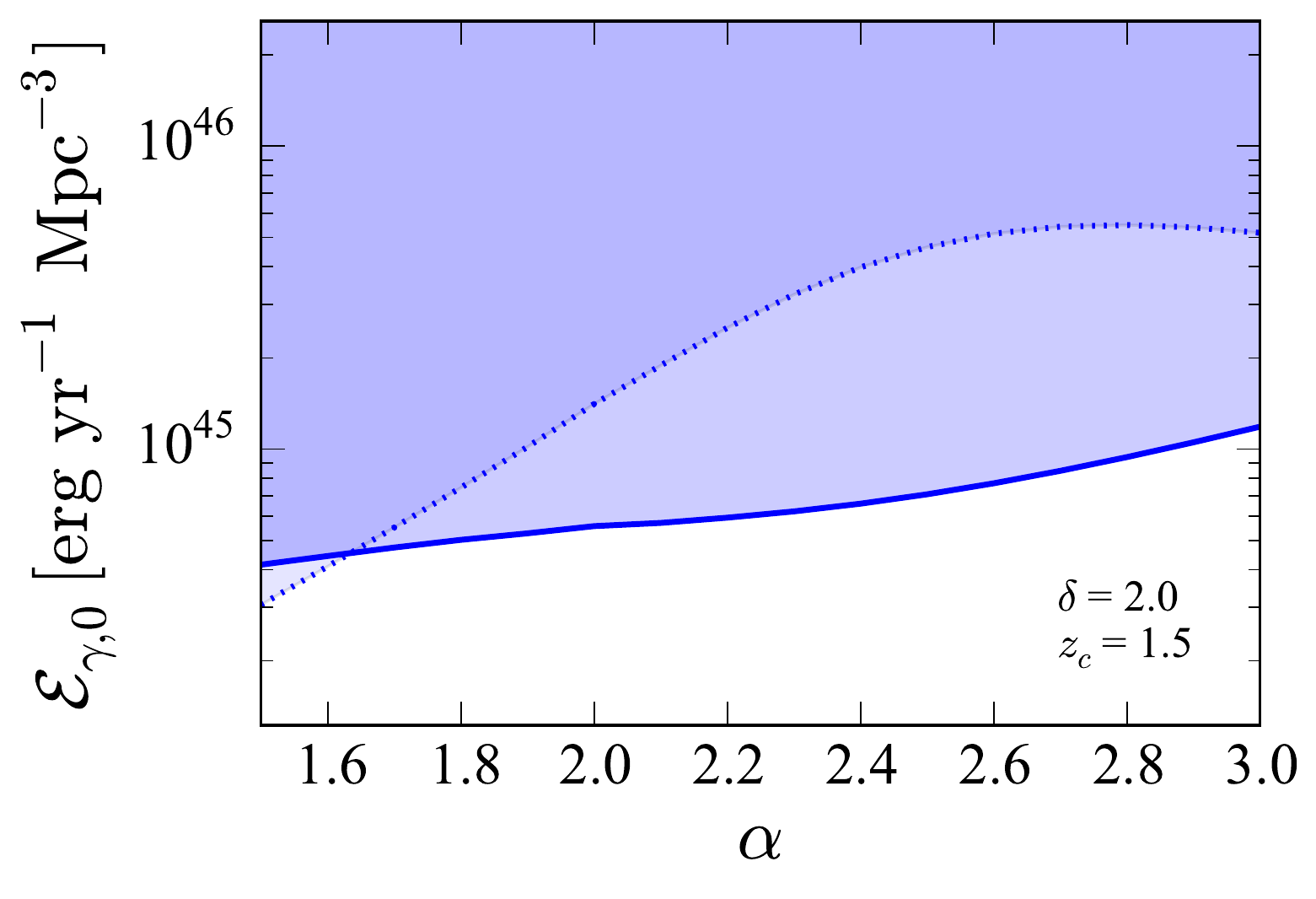}
  \caption{The 95\% CL upper limits on the local gamma-ray luminosity
  density $\mathcal E_{\gamma,0}$, between 100~MeV and 100~GeV, as a
  function of $\alpha$ for $\delta = 2$ and $z_c = 1.5$. The limits due
  to spectrum and cross correlation data are shown as dotted and solid
  curves, respectively.}
  \label{fig:luminosity_density}
 \end{center}
\end{figure}

Figure~\ref{fig:luminosity_density} shows the 95\% CL upper limits on
$\mathcal E_{\gamma,0}$ as a function of $\alpha$, for $\delta = 2$ and
$z_c = 1.5$.
For a wide range of spectral indices, the cross-correlation data provide
constraints more stringent by up to one order of magnitude than the
spectral data.
We also find that the difference is larger for smaller $\delta$, since
the cross correlation constraints are stronger for smaller redshifts,
particularly due to the 2MASS galaxies.
For $\delta \agt 4$, we find that both the spectrum and cross
correlations provide comparable constraints on
$\mathcal{E}_{\gamma,0}$.
The dependence on $z_c$, on the other hand, is significantly weaker as
long as $z_c \ge 1$.

We note that, to be conservative, we did not include
secondary gamma rays that are generated by electromagnetic cascades,
which would improve the spectral constraints~\cite{Murase:2013rfa}.
If the intergalactic magnetic fields are sufficiently weak such that the
cascades do not produce halos or larger diffuse emission
(e.g.,~\cite{Ando:2010rb}), the tomographic constraints will be also 
improved by the same factor.

{\em Constraints on high-energy neutrinos}.---%
If neutrinos are produced by cosmic ray protons via $pp$
interactions, their intensity is  related to that of gamma
rays~\cite{Anchordoqui:2004eu}:
\begin{equation}
\label{eq:nugamma}
 I_\nu(E_\nu) \approx 6 \, I_{\gamma,\textrm{no-EBL}}(E_\gamma) ,
\end{equation}
with $E_\gamma = 2\,E_\nu$.
Here, $I_\nu$ is the neutrino intensity for {\it all} flavors, and
$I_{\gamma, \textrm{no-EBL}}$ is the gamma-ray intensity without EBL
absorption.
Therefore, constraints on $I_\gamma(E_\gamma)$ (or $\mathcal{E}_{\gamma,0}$), for
each set of the parameters $(\alpha, \, \delta, \, z_c)$, can be directly
transformed into those of a neutrino intensity in the TeV--PeV energy
range through Eq.~(\ref{eq:nugamma}).

The top panel of Fig.~\ref{fig:spectrum} shows the 95\% CL upper
limits on $I_\nu(E_\nu)$ from the IGRB spectrum (red dotted) and
the cross-correlation data (red solid) for $\alpha = 2.2$, $\delta
= 2$, and $z_c = 1.5$.
We find that while the IGRB spectrum analysis suggests this particular
model to be compatible with the IceCube data, the tomographic approach
constrains it as a subdominant source.

\begin{figure}
 \begin{center}
  \includegraphics[width=8.5cm]{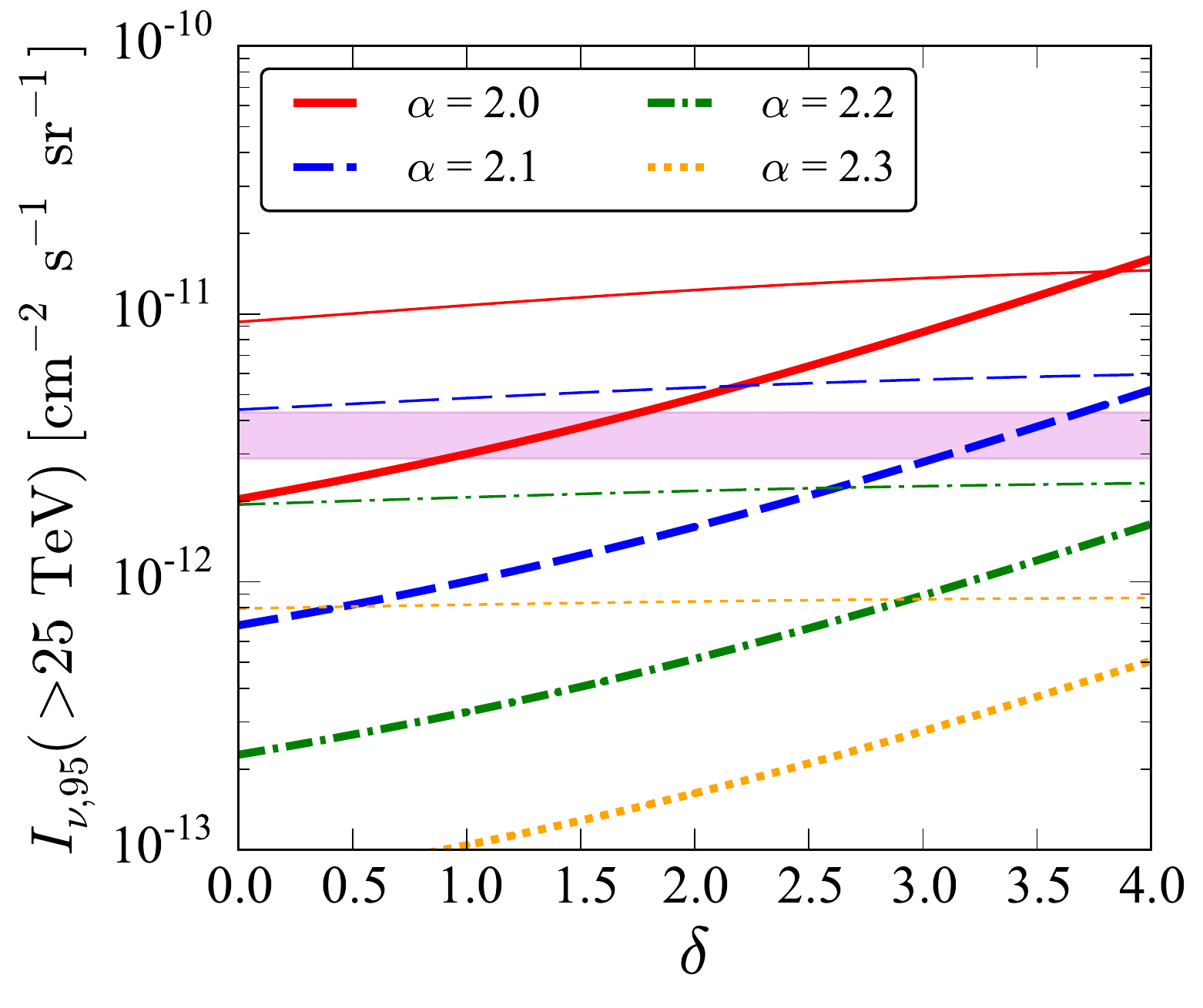}
  \caption{The 95\% CL upper limits on the neutrino intensity integrated
  above 25~TeV as a function of $\delta$ for various values of $\alpha$
  and fixed $z_c = 1.5$. Thick and thin curves show
  the limits due to the tomographic and spectral analyses of the IGRB,
  respectively. The horizontal magenta band shows the 68\% CL interval
  of the best-fit single power-law model for the IceCube neutrino
  data~\cite{Aartsen:2015ita}, corresponding to the neutrino band shown
  in Fig.~\ref{fig:spectrum}.}
  \label{fig:limit_nu}
 \end{center}
\end{figure}

Figure~\ref{fig:limit_nu} shows the dependence of the neutrino intensity
integrated above 25~TeV on $\alpha$ and $\delta$ for a fixed $z_c = 1.5$.
The intensity range preferred by the best-fit single power-law model of
the IceCube data~\cite{Aartsen:2015ita} is also shown for
comparison.
For each model characterized by $(\alpha, \, \delta)$, we show constraints
due to the spectral and tomographic data, as thin and thick curves,
respectively.
Note that the tomographic analysis gives tighter constraints by up to
one order of magnitude with respect to the spectral analysis, especially
for small $\delta$.
In particular, for any source class slowly evolving (e.g., $\delta \alt
3$), even a very hard spectrum such as $E^{-2.1}$ is nearly excluded as
dominant source for the IceCube neutrinos.
Any soft source with $\alpha \agt 2.2$ should contribute much less to
the total neutrino flux than previously expected (e.g.,
Refs.~\cite{Murase:2013rfa, Tamborra:2014xia}).
Models with spectrum as hard as $E^{-2}$, on the other hand, are still
compatible with the IceCube flux level.

{\em Discussion and outlook}.---%
Under the hypothesis that the TeV--PeV IceCube neutrinos are mostly
generated from $pp$ interactions in a single astrophysical source
class (or more classes with similar properties), Fig.~\ref{fig:limit_nu}
implies that a model with $\alpha \approx 2.15$ and $\delta \approx 4$
(for $z_c = 1.5$) can explain most of the neutrino flux.
At the same time, sources of this kind can explain most of the IGRB flux
as well as the measured cross correlations.
We note that in order for such a hard spectrum to be compatible with the
IceCube data, a PeV spectral cutoff is required~\cite{Aartsen:2015ita}
(but data in the northern hemisphere still allow it without a
cutoff~\cite{Aartsen:2015rwa}).
Otherwise, the comparison of the current data set with our results might
suggest a mixed $pp$--$p\gamma$, or even a pure $p\gamma$ origin of the
IceCube neutrino events.

Interestingly, starburst galaxies well satisfy the above conditions for
the $pp$ origin, although efficient cosmic ray confinement needs to be
achieved~\cite{Loeb:2006tw, Tamborra:2014xia, Bartos:2015xpa}.
While direct gamma-ray measurements of the redshift evolution of
star-forming galaxies are not yet available, observations of their
infrared luminosity (or of the star-formation rate) support such steep
evolution.
In particular, the evolution of starburst galaxies 
is characterized by $\delta \gtrsim 4$ up to $z_c \approx
1.5$~\cite{Gruppioni:2013jna}.
Here, we assumed that the local correlation between
infrared and gamma-ray luminosities~\cite{Ackermann:2012vca} holds 
also at high redshifts.

Based on a modeling of resolved gamma-ray sources,
Ref.~\cite{Ajello:2015mfa} argued that about 20--30\% of the IGRB above
100~MeV can be explained by blazars (a subclass of AGNs).
Furthermore, for energies above $\sim$100~GeV, the blazar
contribution can be substantial, explaining most of the IGRB data
and leaving little room for any other source. 
This might point toward an even harder source population with steep
redshift evolution for the neutrinos, which would be, however,
subdominant both in the IGRB flux and cross correlations.
For example, in the case of $\alpha = 2$ and $\delta = 4$, once we
tune the gamma-ray luminosity density to match the level of $\sim$10\%
of the IGRB flux and cross correlations, the same model could explain
most of the neutrino data.

Clusters and groups of galaxies have also been investigated as potential
neutrino sources~\cite{Murase:2013rfa,Zandanel:2014pva}, where cosmic rays, 
generated through large-scale-structure
shocks~\cite{Murase:2008yt,Zandanel:2014pva} or injected by star-forming
galaxies~\cite{Senno:2015tra}, interact with the intracluster medium.
Since the cluster/group number density decreases as a function of
redshift, implying a small value of $\delta$, tomographic constraints
are very stringent.
When considering starbursts or AGNs in clusters/groups, their quick
redshift evolution has to be coupled with the negative one of clusters.
As an example, we calculated that the overall evolution is locally
characterized by $\delta < 2$ that quickly decreases to negative values
for $z \gtrsim 0.5$.
In addition, clusters are largely biased with respect to dark matter
(i.e., $b_{\gamma} \sim 5$ for $10^{15}M_\odot$ and $z =
0$~\cite{Tinker:2010my}), making the tomographic constraints tighter
than those shown in Fig.~\ref{fig:limit_nu}.
Therefore, clusters and groups are disfavored by the cross-correlation
data.

These arguments cannot be applied to $p\gamma$ sources,
such as AGNs~\cite{Stecker:2013fxa, Murase:2014foa, Dermer:2014vaa,
Kalashev:2014vya, Padovani:2015mba}.
This is because the threshold for $p\gamma$ interactions is typically
very high.
It is also argued that such sources may be optically thick for GeV gamma
rays~\cite{Murase:2015xka}.
In any case, it appears difficult that AGNs can be responsible for all 
the IceCube neutrino events.
In fact, Ref.~\cite{Padovani:2015mba} recently suggested that the
diffuse emission from blazars can explain the IceCube neutrino flux at
energies above $\sim$PeV only.

In conclusion, the tomographic method that we apply for the first
time to high-energy neutrinos yields tight constraints on the properties
of any hadronuclear source, providing complementary bounds on their
injection spectral index and redshift evolution.
In particular, we show that only hard spectrum sources with fast
redshift evolution can produce a neutrino flux at the same level as the
IceCube measurement.
The potential relevance of this method in connection with high-energy
neutrinos is expected to quickly increase in the near future, because of
the growing galaxy samples for the cross-correlation analysis, including
cosmic shear measurements that already seem
promising~\cite{Camera:2014rja, Shirasaki:2014noa, Fornengo:2014cya}.

\acknowledgments

We thank Alessandro Cuoco for sharing the cross-correlation data and their
covariance matrices presented in Ref.~\cite{Xia:2015wka}
and for comments on the manuscript, and Kohta Murase for discussions.
This work was supported by the Netherlands Organization for Scientific
Research (NWO) through Vidi (SA and IT) and Veni (FZ) Grants.

\bibliographystyle{apsrev4-1}
\bibliography{refs}

\begin{thebibliography}{69}%
\makeatletter
\providecommand \@ifxundefined [1]{%
 \@ifx{#1\undefined}
}%
\providecommand \@ifnum [1]{%
 \ifnum #1\expandafter \@firstoftwo
 \else \expandafter \@secondoftwo
 \fi
}%
\providecommand \@ifx [1]{%
 \ifx #1\expandafter \@firstoftwo
 \else \expandafter \@secondoftwo
 \fi
}%
\providecommand \natexlab [1]{#1}%
\providecommand \enquote  [1]{``#1''}%
\providecommand \bibnamefont  [1]{#1}%
\providecommand \bibfnamefont [1]{#1}%
\providecommand \citenamefont [1]{#1}%
\providecommand \href@noop [0]{\@secondoftwo}%
\providecommand \href [0]{\begingroup \@sanitize@url \@href}%
\providecommand \@href[1]{\@@startlink{#1}\@@href}%
\providecommand \@@href[1]{\endgroup#1\@@endlink}%
\providecommand \@sanitize@url [0]{\catcode `\\12\catcode `\$12\catcode
  `\&12\catcode `\#12\catcode `\^12\catcode `\_12\catcode `\%12\relax}%
\providecommand \@@startlink[1]{}%
\providecommand \@@endlink[0]{}%
\providecommand \url  [0]{\begingroup\@sanitize@url \@url }%
\providecommand \@url [1]{\endgroup\@href {#1}{\urlprefix }}%
\providecommand \urlprefix  [0]{URL }%
\providecommand \Eprint [0]{\href }%
\providecommand \doibase [0]{http://dx.doi.org/}%
\providecommand \selectlanguage [0]{\@gobble}%
\providecommand \bibinfo  [0]{\@secondoftwo}%
\providecommand \bibfield  [0]{\@secondoftwo}%
\providecommand \translation [1]{[#1]}%
\providecommand \BibitemOpen [0]{}%
\providecommand \bibitemStop [0]{}%
\providecommand \bibitemNoStop [0]{.\EOS\space}%
\providecommand \EOS [0]{\spacefactor3000\relax}%
\providecommand \BibitemShut  [1]{\csname bibitem#1\endcsname}%
\let\auto@bib@innerbib\@empty
\bibitem [{\citenamefont {Aartsen}\ \emph
  {et~al.}(2013{\natexlab{a}})\citenamefont {Aartsen} \emph
  {et~al.}}]{Aartsen:2013bka}%
  \BibitemOpen
  \bibfield  {author} {\bibinfo {author} {\bibfnamefont {M.~G.}\ \bibnamefont
  {Aartsen}} \emph {et~al.} (\bibinfo {collaboration} {IceCube}),\ }\href
  {\doibase 10.1103/PhysRevLett.111.021103} {\bibfield  {journal} {\bibinfo
  {journal} {Phys. Rev. Lett.}\ }\textbf {\bibinfo {volume} {111}},\ \bibinfo
  {pages} {021103} (\bibinfo {year} {2013}{\natexlab{a}})},\ \Eprint
  {http://arxiv.org/abs/1304.5356} {arXiv:1304.5356 [astro-ph.HE]} \BibitemShut
  {NoStop}%
\bibitem [{\citenamefont {Aartsen}\ \emph
  {et~al.}(2013{\natexlab{b}})\citenamefont {Aartsen} \emph
  {et~al.}}]{Aartsen:2013jdh}%
  \BibitemOpen
  \bibfield  {author} {\bibinfo {author} {\bibfnamefont {M.~G.}\ \bibnamefont
  {Aartsen}} \emph {et~al.} (\bibinfo {collaboration} {IceCube}),\ }\href
  {\doibase 10.1126/science.1242856} {\bibfield  {journal} {\bibinfo  {journal}
  {Science}\ }\textbf {\bibinfo {volume} {342}},\ \bibinfo {pages} {1242856}
  (\bibinfo {year} {2013}{\natexlab{b}})},\ \Eprint
  {http://arxiv.org/abs/1311.5238} {arXiv:1311.5238 [astro-ph.HE]} \BibitemShut
  {NoStop}%
\bibitem [{\citenamefont {Aartsen}\ \emph {et~al.}(2014)\citenamefont {Aartsen}
  \emph {et~al.}}]{Aartsen:2014gkd}%
  \BibitemOpen
  \bibfield  {author} {\bibinfo {author} {\bibfnamefont {M.~G.}\ \bibnamefont
  {Aartsen}} \emph {et~al.} (\bibinfo {collaboration} {IceCube}),\ }\href
  {\doibase 10.1103/PhysRevLett.113.101101} {\bibfield  {journal} {\bibinfo
  {journal} {Phys. Rev. Lett.}\ }\textbf {\bibinfo {volume} {113}},\ \bibinfo
  {pages} {101101} (\bibinfo {year} {2014})},\ \Eprint
  {http://arxiv.org/abs/1405.5303} {arXiv:1405.5303 [astro-ph.HE]} \BibitemShut
  {NoStop}%
\bibitem [{\citenamefont {Aartsen}\ \emph
  {et~al.}(2015{\natexlab{a}})\citenamefont {Aartsen} \emph
  {et~al.}}]{Aartsen:2014muf}%
  \BibitemOpen
  \bibfield  {author} {\bibinfo {author} {\bibfnamefont {M.~G.}\ \bibnamefont
  {Aartsen}} \emph {et~al.} (\bibinfo {collaboration} {IceCube}),\ }\href
  {\doibase 10.1103/PhysRevD.91.022001} {\bibfield  {journal} {\bibinfo
  {journal} {Phys. Rev. D}\ }\textbf {\bibinfo {volume} {91}},\ \bibinfo
  {pages} {022001} (\bibinfo {year} {2015}{\natexlab{a}})},\ \Eprint
  {http://arxiv.org/abs/1410.1749} {arXiv:1410.1749 [astro-ph.HE]} \BibitemShut
  {NoStop}%
\bibitem [{\citenamefont {Aartsen}\ \emph
  {et~al.}(2015{\natexlab{b}})\citenamefont {Aartsen} \emph
  {et~al.}}]{Aartsen:2015ita}%
  \BibitemOpen
  \bibfield  {author} {\bibinfo {author} {\bibfnamefont {M.~G.}\ \bibnamefont
  {Aartsen}} \emph {et~al.} (\bibinfo {collaboration} {IceCube}),\ }\href
  {\doibase 10.1088/0004-637X/809/1/98} {\bibfield  {journal} {\bibinfo
  {journal} {Astrophys. J.}\ }\textbf {\bibinfo {volume} {809}},\ \bibinfo
  {pages} {98} (\bibinfo {year} {2015}{\natexlab{b}})},\ \Eprint
  {http://arxiv.org/abs/1507.03991} {arXiv:1507.03991 [astro-ph.HE]}
  \BibitemShut {NoStop}%
\bibitem [{\citenamefont {Aartsen}\ \emph
  {et~al.}(2015{\natexlab{c}})\citenamefont {Aartsen} \emph
  {et~al.}}]{Aartsen:2015rwa}%
  \BibitemOpen
  \bibfield  {author} {\bibinfo {author} {\bibfnamefont {M.~G.}\ \bibnamefont
  {Aartsen}} \emph {et~al.} (\bibinfo {collaboration} {IceCube}),\ }\href
  {\doibase 10.1103/PhysRevLett.115.081102} {\bibfield  {journal} {\bibinfo
  {journal} {Phys. Rev. Lett.}\ }\textbf {\bibinfo {volume} {115}},\ \bibinfo
  {pages} {081102} (\bibinfo {year} {2015}{\natexlab{c}})},\ \Eprint
  {http://arxiv.org/abs/1507.04005} {arXiv:1507.04005 [astro-ph.HE]}
  \BibitemShut {NoStop}%
\bibitem [{\citenamefont {Anchordoqui}\ \emph
  {et~al.}(2014{\natexlab{a}})\citenamefont {Anchordoqui} \emph
  {et~al.}}]{Anchordoqui:2013dnh}%
  \BibitemOpen
  \bibfield  {author} {\bibinfo {author} {\bibfnamefont {L.~A.}\ \bibnamefont
  {Anchordoqui}} \emph {et~al.},\ }\href {\doibase 10.1016/j.jheap.2014.01.001}
  {\bibfield  {journal} {\bibinfo  {journal} {JHEAp}\ }\textbf {\bibinfo
  {volume} {1-2}},\ \bibinfo {pages} {1} (\bibinfo {year}
  {2014}{\natexlab{a}})},\ \Eprint {http://arxiv.org/abs/1312.6587}
  {arXiv:1312.6587 [astro-ph.HE]} \BibitemShut {NoStop}%
\bibitem [{\citenamefont {Waxman}(2013)}]{Waxman:2013zda}%
  \BibitemOpen
  \bibfield  {author} {\bibinfo {author} {\bibfnamefont {E.}~\bibnamefont
  {Waxman}},\ }in\ \href
  {http://inspirehep.net/record/1266927/files/arXiv:1312.0558.pdf} {\emph
  {\bibinfo {booktitle} {{Rencontres du Vietnam: Windows on the Universe Quy
  Nhon, Binh Dinh, Vietnam, August 11-17, 2013}}}}\ (\bibinfo {year} {2013})\
  \Eprint {http://arxiv.org/abs/1312.0558} {arXiv:1312.0558 [astro-ph.HE]}
  \BibitemShut {NoStop}%
\bibitem [{\citenamefont {Stecker}\ \emph {et~al.}(1991)\citenamefont
  {Stecker}, \citenamefont {Done}, \citenamefont {Salamon},\ and\ \citenamefont
  {Sommers}}]{Stecker:1991vm}%
  \BibitemOpen
  \bibfield  {author} {\bibinfo {author} {\bibfnamefont {F.~W.}\ \bibnamefont
  {Stecker}}, \bibinfo {author} {\bibfnamefont {C.}~\bibnamefont {Done}},
  \bibinfo {author} {\bibfnamefont {M.~H.}\ \bibnamefont {Salamon}}, \ and\
  \bibinfo {author} {\bibfnamefont {P.}~\bibnamefont {Sommers}},\ }\href
  {\doibase 10.1103/PhysRevLett.66.2697} {\bibfield  {journal} {\bibinfo
  {journal} {Phys. Rev. Lett.}\ }\textbf {\bibinfo {volume} {66}},\ \bibinfo
  {pages} {2697} (\bibinfo {year} {1991})},\ \bibinfo {note} {[Erratum: Phys.
  Rev. Lett.69,2738(1992)]}\BibitemShut {NoStop}%
\bibitem [{\citenamefont {Stecker}(2005)}]{Stecker:2005hn}%
  \BibitemOpen
  \bibfield  {author} {\bibinfo {author} {\bibfnamefont {F.~W.}\ \bibnamefont
  {Stecker}},\ }\href {\doibase 10.1103/PhysRevD.72.107301} {\bibfield
  {journal} {\bibinfo  {journal} {Phys. Rev. D}\ }\textbf {\bibinfo {volume}
  {72}},\ \bibinfo {pages} {107301} (\bibinfo {year} {2005})},\ \Eprint
  {http://arxiv.org/abs/astro-ph/0510537} {arXiv:astro-ph/0510537 [astro-ph]}
  \BibitemShut {NoStop}%
\bibitem [{\citenamefont {Kalashev}\ \emph {et~al.}(2013)\citenamefont
  {Kalashev}, \citenamefont {Kusenko},\ and\ \citenamefont
  {Essey}}]{Kalashev:2013vba}%
  \BibitemOpen
  \bibfield  {author} {\bibinfo {author} {\bibfnamefont {O.~E.}\ \bibnamefont
  {Kalashev}}, \bibinfo {author} {\bibfnamefont {A.}~\bibnamefont {Kusenko}}, \
  and\ \bibinfo {author} {\bibfnamefont {W.}~\bibnamefont {Essey}},\ }\href
  {\doibase 10.1103/PhysRevLett.111.041103} {\bibfield  {journal} {\bibinfo
  {journal} {Phys. Rev. Lett.}\ }\textbf {\bibinfo {volume} {111}},\ \bibinfo
  {pages} {041103} (\bibinfo {year} {2013})},\ \Eprint
  {http://arxiv.org/abs/1303.0300} {arXiv:1303.0300 [astro-ph.HE]} \BibitemShut
  {NoStop}%
\bibitem [{\citenamefont {Stecker}(2013)}]{Stecker:2013fxa}%
  \BibitemOpen
  \bibfield  {author} {\bibinfo {author} {\bibfnamefont {F.~W.}\ \bibnamefont
  {Stecker}},\ }\href {\doibase 10.1103/PhysRevD.88.047301} {\bibfield
  {journal} {\bibinfo  {journal} {Phys. Rev. D}\ }\textbf {\bibinfo {volume}
  {88}},\ \bibinfo {pages} {047301} (\bibinfo {year} {2013})},\ \Eprint
  {http://arxiv.org/abs/1305.7404} {arXiv:1305.7404 [astro-ph.HE]} \BibitemShut
  {NoStop}%
\bibitem [{\citenamefont {Murase}\ \emph {et~al.}(2014)\citenamefont {Murase},
  \citenamefont {Inoue},\ and\ \citenamefont {Dermer}}]{Murase:2014foa}%
  \BibitemOpen
  \bibfield  {author} {\bibinfo {author} {\bibfnamefont {K.}~\bibnamefont
  {Murase}}, \bibinfo {author} {\bibfnamefont {Y.}~\bibnamefont {Inoue}}, \
  and\ \bibinfo {author} {\bibfnamefont {C.~D.}\ \bibnamefont {Dermer}},\
  }\href {\doibase 10.1103/PhysRevD.90.023007} {\bibfield  {journal} {\bibinfo
  {journal} {Phys. Rev. D}\ }\textbf {\bibinfo {volume} {90}},\ \bibinfo
  {pages} {023007} (\bibinfo {year} {2014})},\ \Eprint
  {http://arxiv.org/abs/1403.4089} {arXiv:1403.4089 [astro-ph.HE]} \BibitemShut
  {NoStop}%
\bibitem [{\citenamefont {Kimura}\ \emph {et~al.}(2015)\citenamefont {Kimura},
  \citenamefont {Murase},\ and\ \citenamefont {Toma}}]{Kimura:2014jba}%
  \BibitemOpen
  \bibfield  {author} {\bibinfo {author} {\bibfnamefont {S.~S.}\ \bibnamefont
  {Kimura}}, \bibinfo {author} {\bibfnamefont {K.}~\bibnamefont {Murase}}, \
  and\ \bibinfo {author} {\bibfnamefont {K.}~\bibnamefont {Toma}},\ }\href
  {\doibase 10.1088/0004-637X/806/2/159} {\bibfield  {journal} {\bibinfo
  {journal} {Astrophys. J.}\ }\textbf {\bibinfo {volume} {806}},\ \bibinfo
  {pages} {159} (\bibinfo {year} {2015})},\ \Eprint
  {http://arxiv.org/abs/1411.3588} {arXiv:1411.3588 [astro-ph.HE]} \BibitemShut
  {NoStop}%
\bibitem [{\citenamefont {Dermer}\ \emph {et~al.}(2014)\citenamefont {Dermer},
  \citenamefont {Murase},\ and\ \citenamefont {Inoue}}]{Dermer:2014vaa}%
  \BibitemOpen
  \bibfield  {author} {\bibinfo {author} {\bibfnamefont {C.~D.}\ \bibnamefont
  {Dermer}}, \bibinfo {author} {\bibfnamefont {K.}~\bibnamefont {Murase}}, \
  and\ \bibinfo {author} {\bibfnamefont {Y.}~\bibnamefont {Inoue}},\ }\href
  {\doibase 10.1016/j.jheap.2014.09.001} {\bibfield  {journal} {\bibinfo
  {journal} {JHEAp}\ }\textbf {\bibinfo {volume} {3-4}},\ \bibinfo {pages} {29}
  (\bibinfo {year} {2014})},\ \Eprint {http://arxiv.org/abs/1406.2633}
  {arXiv:1406.2633 [astro-ph.HE]} \BibitemShut {NoStop}%
\bibitem [{\citenamefont {Kalashev}\ \emph {et~al.}(2014)\citenamefont
  {Kalashev}, \citenamefont {Semikoz},\ and\ \citenamefont
  {Tkachev}}]{Kalashev:2014vya}%
  \BibitemOpen
  \bibfield  {author} {\bibinfo {author} {\bibfnamefont {O.}~\bibnamefont
  {Kalashev}}, \bibinfo {author} {\bibfnamefont {D.}~\bibnamefont {Semikoz}}, \
  and\ \bibinfo {author} {\bibfnamefont {I.}~\bibnamefont {Tkachev}},\
  }\href@noop {} {\  (\bibinfo {year} {2014})},\ \Eprint
  {http://arxiv.org/abs/1410.8124} {arXiv:1410.8124 [astro-ph.HE]} \BibitemShut
  {NoStop}%
\bibitem [{\citenamefont {Khiali}\ and\ \citenamefont
  {Pino}(2015)}]{Khiali:2015tfa}%
  \BibitemOpen
  \bibfield  {author} {\bibinfo {author} {\bibfnamefont {B.}~\bibnamefont
  {Khiali}}\ and\ \bibinfo {author} {\bibfnamefont {E.~M. d. G.~D.}\
  \bibnamefont {Pino}},\ }\href@noop {} {\  (\bibinfo {year} {2015})},\ \Eprint
  {http://arxiv.org/abs/1506.01063} {arXiv:1506.01063 [astro-ph.HE]}
  \BibitemShut {NoStop}%
\bibitem [{\citenamefont {{Padovani}}\ \emph {et~al.}(2015)\citenamefont
  {{Padovani}}, \citenamefont {{Petropoulou}}, \citenamefont {{Giommi}},\ and\
  \citenamefont {{Resconi}}}]{Padovani:2015mba}%
  \BibitemOpen
  \bibfield  {author} {\bibinfo {author} {\bibfnamefont {P.}~\bibnamefont
  {{Padovani}}}, \bibinfo {author} {\bibfnamefont {M.}~\bibnamefont
  {{Petropoulou}}}, \bibinfo {author} {\bibfnamefont {P.}~\bibnamefont
  {{Giommi}}}, \ and\ \bibinfo {author} {\bibfnamefont {E.}~\bibnamefont
  {{Resconi}}},\ }\href {\doibase 10.1093/mnras/stv1467} {\bibfield  {journal}
  {\bibinfo  {journal} {Mon. Not. R. Astron. Soc.}\ }\textbf {\bibinfo {volume}
  {452}},\ \bibinfo {pages} {1877} (\bibinfo {year} {2015})},\ \Eprint
  {http://arxiv.org/abs/1506.09135} {arXiv:1506.09135 [astro-ph.HE]}
  \BibitemShut {NoStop}%
\bibitem [{\citenamefont {Loeb}\ and\ \citenamefont
  {Waxman}(2006)}]{Loeb:2006tw}%
  \BibitemOpen
  \bibfield  {author} {\bibinfo {author} {\bibfnamefont {A.}~\bibnamefont
  {Loeb}}\ and\ \bibinfo {author} {\bibfnamefont {E.}~\bibnamefont {Waxman}},\
  }\href {\doibase 10.1088/1475-7516/2006/05/003} {\bibfield  {journal}
  {\bibinfo  {journal} {JCAP}\ }\textbf {\bibinfo {volume} {0605}},\ \bibinfo
  {pages} {003} (\bibinfo {year} {2006})},\ \Eprint
  {http://arxiv.org/abs/astro-ph/0601695} {arXiv:astro-ph/0601695 [astro-ph]}
  \BibitemShut {NoStop}%
\bibitem [{\citenamefont {Thompson}\ \emph {et~al.}(2006)\citenamefont
  {Thompson}, \citenamefont {Quataert}, \citenamefont {Waxman},\ and\
  \citenamefont {Loeb}}]{Thompson:2006np}%
  \BibitemOpen
  \bibfield  {author} {\bibinfo {author} {\bibfnamefont {T.~A.}\ \bibnamefont
  {Thompson}}, \bibinfo {author} {\bibfnamefont {E.}~\bibnamefont {Quataert}},
  \bibinfo {author} {\bibfnamefont {E.}~\bibnamefont {Waxman}}, \ and\ \bibinfo
  {author} {\bibfnamefont {A.}~\bibnamefont {Loeb}},\ }\href@noop {} {\
  (\bibinfo {year} {2006})},\ \Eprint {http://arxiv.org/abs/astro-ph/0608699}
  {arXiv:astro-ph/0608699 [astro-ph]} \BibitemShut {NoStop}%
\bibitem [{\citenamefont {He}\ \emph {et~al.}(2013)\citenamefont {He},
  \citenamefont {Wang}, \citenamefont {Fan}, \citenamefont {Liu},\ and\
  \citenamefont {Wei}}]{He:2013cqa}%
  \BibitemOpen
  \bibfield  {author} {\bibinfo {author} {\bibfnamefont {H.-N.}\ \bibnamefont
  {He}}, \bibinfo {author} {\bibfnamefont {T.}~\bibnamefont {Wang}}, \bibinfo
  {author} {\bibfnamefont {Y.-Z.}\ \bibnamefont {Fan}}, \bibinfo {author}
  {\bibfnamefont {S.-M.}\ \bibnamefont {Liu}}, \ and\ \bibinfo {author}
  {\bibfnamefont {D.-M.}\ \bibnamefont {Wei}},\ }\href {\doibase
  10.1103/PhysRevD.87.063011} {\bibfield  {journal} {\bibinfo  {journal} {Phys.
  Rev. D}\ }\textbf {\bibinfo {volume} {87}},\ \bibinfo {pages} {063011}
  (\bibinfo {year} {2013})},\ \Eprint {http://arxiv.org/abs/1303.1253}
  {arXiv:1303.1253 [astro-ph.HE]} \BibitemShut {NoStop}%
\bibitem [{\citenamefont {Liu}\ \emph {et~al.}(2014)\citenamefont {Liu},
  \citenamefont {Wang}, \citenamefont {Inoue}, \citenamefont {Crocker},\ and\
  \citenamefont {Aharonian}}]{Liu:2013wia}%
  \BibitemOpen
  \bibfield  {author} {\bibinfo {author} {\bibfnamefont {R.-Y.}\ \bibnamefont
  {Liu}}, \bibinfo {author} {\bibfnamefont {X.-Y.}\ \bibnamefont {Wang}},
  \bibinfo {author} {\bibfnamefont {S.}~\bibnamefont {Inoue}}, \bibinfo
  {author} {\bibfnamefont {R.}~\bibnamefont {Crocker}}, \ and\ \bibinfo
  {author} {\bibfnamefont {F.}~\bibnamefont {Aharonian}},\ }\href {\doibase
  10.1103/PhysRevD.89.083004} {\bibfield  {journal} {\bibinfo  {journal} {Phys.
  Rev. D}\ }\textbf {\bibinfo {volume} {89}},\ \bibinfo {pages} {083004}
  (\bibinfo {year} {2014})},\ \Eprint {http://arxiv.org/abs/1310.1263}
  {arXiv:1310.1263 [astro-ph.HE]} \BibitemShut {NoStop}%
\bibitem [{\citenamefont {Chang}\ and\ \citenamefont
  {Wang}(2014)}]{Chang:2014hua}%
  \BibitemOpen
  \bibfield  {author} {\bibinfo {author} {\bibfnamefont {X.-C.}\ \bibnamefont
  {Chang}}\ and\ \bibinfo {author} {\bibfnamefont {X.-Y.}\ \bibnamefont
  {Wang}},\ }\href {\doibase 10.1088/0004-637X/793/2/131} {\bibfield  {journal}
  {\bibinfo  {journal} {Astrophys. J.}\ }\textbf {\bibinfo {volume} {793}},\
  \bibinfo {pages} {131} (\bibinfo {year} {2014})},\ \Eprint
  {http://arxiv.org/abs/1406.1099} {arXiv:1406.1099 [astro-ph.HE]} \BibitemShut
  {NoStop}%
\bibitem [{\citenamefont {Tamborra}\ \emph {et~al.}(2014)\citenamefont
  {Tamborra}, \citenamefont {Ando},\ and\ \citenamefont
  {Murase}}]{Tamborra:2014xia}%
  \BibitemOpen
  \bibfield  {author} {\bibinfo {author} {\bibfnamefont {I.}~\bibnamefont
  {Tamborra}}, \bibinfo {author} {\bibfnamefont {S.}~\bibnamefont {Ando}}, \
  and\ \bibinfo {author} {\bibfnamefont {K.}~\bibnamefont {Murase}},\ }\href
  {\doibase 10.1088/1475-7516/2014/09/043} {\bibfield  {journal} {\bibinfo
  {journal} {JCAP}\ }\textbf {\bibinfo {volume} {1409}},\ \bibinfo {pages}
  {043} (\bibinfo {year} {2014})},\ \Eprint {http://arxiv.org/abs/1404.1189}
  {arXiv:1404.1189 [astro-ph.HE]} \BibitemShut {NoStop}%
\bibitem [{\citenamefont {Anchordoqui}\ \emph
  {et~al.}(2014{\natexlab{b}})\citenamefont {Anchordoqui}, \citenamefont
  {Paul}, \citenamefont {da~Silva}, \citenamefont {Torres},\ and\ \citenamefont
  {Vlcek}}]{Anchordoqui:2014yva}%
  \BibitemOpen
  \bibfield  {author} {\bibinfo {author} {\bibfnamefont {L.~A.}\ \bibnamefont
  {Anchordoqui}}, \bibinfo {author} {\bibfnamefont {T.~C.}\ \bibnamefont
  {Paul}}, \bibinfo {author} {\bibfnamefont {L.~H.~M.}\ \bibnamefont
  {da~Silva}}, \bibinfo {author} {\bibfnamefont {D.~F.}\ \bibnamefont
  {Torres}}, \ and\ \bibinfo {author} {\bibfnamefont {B.~J.}\ \bibnamefont
  {Vlcek}},\ }\href {\doibase 10.1103/PhysRevD.89.127304} {\bibfield  {journal}
  {\bibinfo  {journal} {Phys. Rev. D}\ }\textbf {\bibinfo {volume} {89}},\
  \bibinfo {pages} {127304} (\bibinfo {year} {2014}{\natexlab{b}})},\ \Eprint
  {http://arxiv.org/abs/1405.7648} {arXiv:1405.7648 [astro-ph.HE]} \BibitemShut
  {NoStop}%
\bibitem [{\citenamefont {Chakraborty}\ and\ \citenamefont
  {Izaguirre}(2015)}]{Chakraborty:2015sta}%
  \BibitemOpen
  \bibfield  {author} {\bibinfo {author} {\bibfnamefont {S.}~\bibnamefont
  {Chakraborty}}\ and\ \bibinfo {author} {\bibfnamefont {I.}~\bibnamefont
  {Izaguirre}},\ }\href {\doibase 10.1016/j.physletb.2015.04.032} {\bibfield
  {journal} {\bibinfo  {journal} {Phys. Lett. B}\ }\textbf {\bibinfo {volume}
  {745}},\ \bibinfo {pages} {35} (\bibinfo {year} {2015})},\ \Eprint
  {http://arxiv.org/abs/1501.02615} {arXiv:1501.02615 [hep-ph]} \BibitemShut
  {NoStop}%
\bibitem [{\citenamefont {Senno}\ \emph {et~al.}(2015)\citenamefont {Senno},
  \citenamefont {M{\'e}sz{\'a}ros}, \citenamefont {Murase}, \citenamefont
  {Baerwald},\ and\ \citenamefont {Rees}}]{Senno:2015tra}%
  \BibitemOpen
  \bibfield  {author} {\bibinfo {author} {\bibfnamefont {N.}~\bibnamefont
  {Senno}}, \bibinfo {author} {\bibfnamefont {P.}~\bibnamefont
  {M{\'e}sz{\'a}ros}}, \bibinfo {author} {\bibfnamefont {K.}~\bibnamefont
  {Murase}}, \bibinfo {author} {\bibfnamefont {P.}~\bibnamefont {Baerwald}}, \
  and\ \bibinfo {author} {\bibfnamefont {M.~J.}\ \bibnamefont {Rees}},\ }\href
  {\doibase 10.1088/0004-637X/806/1/24} {\bibfield  {journal} {\bibinfo
  {journal} {Astrophys. J.}\ }\textbf {\bibinfo {volume} {806}},\ \bibinfo
  {pages} {24} (\bibinfo {year} {2015})},\ \Eprint
  {http://arxiv.org/abs/1501.04934} {arXiv:1501.04934 [astro-ph.HE]}
  \BibitemShut {NoStop}%
\bibitem [{\citenamefont {Bartos}\ and\ \citenamefont
  {Marka}(2015)}]{Bartos:2015xpa}%
  \BibitemOpen
  \bibfield  {author} {\bibinfo {author} {\bibfnamefont {I.}~\bibnamefont
  {Bartos}}\ and\ \bibinfo {author} {\bibfnamefont {S.}~\bibnamefont {Marka}},\
  }\href@noop {} {\  (\bibinfo {year} {2015})},\ \Eprint
  {http://arxiv.org/abs/1509.00983} {arXiv:1509.00983 [astro-ph.HE]}
  \BibitemShut {NoStop}%
\bibitem [{\citenamefont {Waxman}\ and\ \citenamefont
  {Bahcall}(1999)}]{Waxman:1998yy}%
  \BibitemOpen
  \bibfield  {author} {\bibinfo {author} {\bibfnamefont {E.}~\bibnamefont
  {Waxman}}\ and\ \bibinfo {author} {\bibfnamefont {J.~N.}\ \bibnamefont
  {Bahcall}},\ }\href {\doibase 10.1103/PhysRevD.59.023002} {\bibfield
  {journal} {\bibinfo  {journal} {Phys. Rev. D}\ }\textbf {\bibinfo {volume}
  {59}},\ \bibinfo {pages} {023002} (\bibinfo {year} {1999})},\ \Eprint
  {http://arxiv.org/abs/hep-ph/9807282} {arXiv:hep-ph/9807282 [hep-ph]}
  \BibitemShut {NoStop}%
\bibitem [{\citenamefont {M\'esz\'aros}\ and\ \citenamefont
  {Waxman}(2001)}]{Meszaros:2001ms}%
  \BibitemOpen
  \bibfield  {author} {\bibinfo {author} {\bibfnamefont {P.}~\bibnamefont
  {M\'esz\'aros}}\ and\ \bibinfo {author} {\bibfnamefont {E.}~\bibnamefont
  {Waxman}},\ }\href {\doibase 10.1103/PhysRevLett.87.171102} {\bibfield
  {journal} {\bibinfo  {journal} {Phys. Rev. Lett.}\ }\textbf {\bibinfo
  {volume} {87}},\ \bibinfo {pages} {171102} (\bibinfo {year} {2001})},\
  \Eprint {http://arxiv.org/abs/astro-ph/0103275} {arXiv:astro-ph/0103275
  [astro-ph]} \BibitemShut {NoStop}%
\bibitem [{\citenamefont {Murase}\ and\ \citenamefont
  {Nagataki}(2006)}]{Murase:2005hy}%
  \BibitemOpen
  \bibfield  {author} {\bibinfo {author} {\bibfnamefont {K.}~\bibnamefont
  {Murase}}\ and\ \bibinfo {author} {\bibfnamefont {S.}~\bibnamefont
  {Nagataki}},\ }\href {\doibase 10.1103/PhysRevD.73.063002} {\bibfield
  {journal} {\bibinfo  {journal} {Phys. Rev. D}\ }\textbf {\bibinfo {volume}
  {73}},\ \bibinfo {pages} {063002} (\bibinfo {year} {2006})},\ \Eprint
  {http://arxiv.org/abs/astro-ph/0512275} {arXiv:astro-ph/0512275 [astro-ph]}
  \BibitemShut {NoStop}%
\bibitem [{\citenamefont {Gupta}\ and\ \citenamefont
  {Zhang}(2007)}]{Gupta:2006jm}%
  \BibitemOpen
  \bibfield  {author} {\bibinfo {author} {\bibfnamefont {N.}~\bibnamefont
  {Gupta}}\ and\ \bibinfo {author} {\bibfnamefont {B.}~\bibnamefont {Zhang}},\
  }\href {\doibase 10.1016/j.astropartphys.2007.01.004} {\bibfield  {journal}
  {\bibinfo  {journal} {Astropart. Phys.}\ }\textbf {\bibinfo {volume} {27}},\
  \bibinfo {pages} {386} (\bibinfo {year} {2007})},\ \Eprint
  {http://arxiv.org/abs/astro-ph/0606744} {arXiv:astro-ph/0606744 [astro-ph]}
  \BibitemShut {NoStop}%
\bibitem [{\citenamefont {Murase}\ \emph {et~al.}(2006)\citenamefont {Murase},
  \citenamefont {Ioka}, \citenamefont {Nagataki},\ and\ \citenamefont
  {Nakamura}}]{Murase:2006mm}%
  \BibitemOpen
  \bibfield  {author} {\bibinfo {author} {\bibfnamefont {K.}~\bibnamefont
  {Murase}}, \bibinfo {author} {\bibfnamefont {K.}~\bibnamefont {Ioka}},
  \bibinfo {author} {\bibfnamefont {S.}~\bibnamefont {Nagataki}}, \ and\
  \bibinfo {author} {\bibfnamefont {T.}~\bibnamefont {Nakamura}},\ }\href
  {\doibase 10.1086/509323} {\bibfield  {journal} {\bibinfo  {journal}
  {Astrophys. J.}\ }\textbf {\bibinfo {volume} {651}},\ \bibinfo {pages} {L5}
  (\bibinfo {year} {2006})},\ \Eprint {http://arxiv.org/abs/astro-ph/0607104}
  {arXiv:astro-ph/0607104 [astro-ph]} \BibitemShut {NoStop}%
\bibitem [{\citenamefont {Liu}\ and\ \citenamefont {Wang}(2013)}]{Liu:2012pf}%
  \BibitemOpen
  \bibfield  {author} {\bibinfo {author} {\bibfnamefont {R.-Y.}\ \bibnamefont
  {Liu}}\ and\ \bibinfo {author} {\bibfnamefont {X.-Y.}\ \bibnamefont {Wang}},\
  }\href {\doibase 10.1088/0004-637X/766/2/73} {\bibfield  {journal} {\bibinfo
  {journal} {Astrophys. J.}\ }\textbf {\bibinfo {volume} {766}},\ \bibinfo
  {pages} {73} (\bibinfo {year} {2013})},\ \Eprint
  {http://arxiv.org/abs/1212.1260} {arXiv:1212.1260 [astro-ph.HE]} \BibitemShut
  {NoStop}%
\bibitem [{\citenamefont {Murase}\ and\ \citenamefont
  {Ioka}(2013)}]{Murase:2013ffa}%
  \BibitemOpen
  \bibfield  {author} {\bibinfo {author} {\bibfnamefont {K.}~\bibnamefont
  {Murase}}\ and\ \bibinfo {author} {\bibfnamefont {K.}~\bibnamefont {Ioka}},\
  }\href {\doibase 10.1103/PhysRevLett.111.121102} {\bibfield  {journal}
  {\bibinfo  {journal} {Phys. Rev. Lett.}\ }\textbf {\bibinfo {volume} {111}},\
  \bibinfo {pages} {121102} (\bibinfo {year} {2013})},\ \Eprint
  {http://arxiv.org/abs/1306.2274} {arXiv:1306.2274 [astro-ph.HE]} \BibitemShut
  {NoStop}%
\bibitem [{\citenamefont {Tamborra}\ and\ \citenamefont
  {Ando}(2015)}]{Tamborra:2015qza}%
  \BibitemOpen
  \bibfield  {author} {\bibinfo {author} {\bibfnamefont {I.}~\bibnamefont
  {Tamborra}}\ and\ \bibinfo {author} {\bibfnamefont {S.}~\bibnamefont
  {Ando}},\ }\href@noop {} {\  (\bibinfo {year} {2015})},\ \Eprint
  {http://arxiv.org/abs/1504.00107} {arXiv:1504.00107 [astro-ph.HE]}
  \BibitemShut {NoStop}%
\bibitem [{\citenamefont {Murase}\ \emph {et~al.}(2008)\citenamefont {Murase},
  \citenamefont {Inoue},\ and\ \citenamefont {Nagataki}}]{Murase:2008yt}%
  \BibitemOpen
  \bibfield  {author} {\bibinfo {author} {\bibfnamefont {K.}~\bibnamefont
  {Murase}}, \bibinfo {author} {\bibfnamefont {S.}~\bibnamefont {Inoue}}, \
  and\ \bibinfo {author} {\bibfnamefont {S.}~\bibnamefont {Nagataki}},\ }\href
  {\doibase 10.1086/595882} {\bibfield  {journal} {\bibinfo  {journal}
  {Astrophys. J.}\ }\textbf {\bibinfo {volume} {689}},\ \bibinfo {pages} {L105}
  (\bibinfo {year} {2008})},\ \Eprint {http://arxiv.org/abs/0805.0104}
  {arXiv:0805.0104 [astro-ph]} \BibitemShut {NoStop}%
\bibitem [{\citenamefont {Kotera}\ \emph {et~al.}(2009)\citenamefont {Kotera},
  \citenamefont {Allard}, \citenamefont {Murase}, \citenamefont {Aoi},
  \citenamefont {Dubois}, \citenamefont {Pierog},\ and\ \citenamefont
  {Nagataki}}]{Kotera:2009ms}%
  \BibitemOpen
  \bibfield  {author} {\bibinfo {author} {\bibfnamefont {K.}~\bibnamefont
  {Kotera}}, \bibinfo {author} {\bibfnamefont {D.}~\bibnamefont {Allard}},
  \bibinfo {author} {\bibfnamefont {K.}~\bibnamefont {Murase}}, \bibinfo
  {author} {\bibfnamefont {J.}~\bibnamefont {Aoi}}, \bibinfo {author}
  {\bibfnamefont {Y.}~\bibnamefont {Dubois}}, \bibinfo {author} {\bibfnamefont
  {T.}~\bibnamefont {Pierog}}, \ and\ \bibinfo {author} {\bibfnamefont
  {S.}~\bibnamefont {Nagataki}},\ }\href {\doibase 10.1088/0004-637X/707/1/370}
  {\bibfield  {journal} {\bibinfo  {journal} {Astrophys. J.}\ }\textbf
  {\bibinfo {volume} {707}},\ \bibinfo {pages} {370} (\bibinfo {year}
  {2009})},\ \Eprint {http://arxiv.org/abs/0907.2433} {arXiv:0907.2433
  [astro-ph.HE]} \BibitemShut {NoStop}%
\bibitem [{\citenamefont {Murase}\ and\ \citenamefont
  {Beacom}(2013)}]{Murase:2012rd}%
  \BibitemOpen
  \bibfield  {author} {\bibinfo {author} {\bibfnamefont {K.}~\bibnamefont
  {Murase}}\ and\ \bibinfo {author} {\bibfnamefont {J.~F.}\ \bibnamefont
  {Beacom}},\ }\href {\doibase 10.1088/1475-7516/2013/02/028} {\bibfield
  {journal} {\bibinfo  {journal} {JCAP}\ }\textbf {\bibinfo {volume} {1302}},\
  \bibinfo {pages} {028} (\bibinfo {year} {2013})},\ \Eprint
  {http://arxiv.org/abs/1209.0225} {arXiv:1209.0225 [astro-ph.HE]} \BibitemShut
  {NoStop}%
\bibitem [{\citenamefont {Zandanel}\ \emph {et~al.}(2015)\citenamefont
  {Zandanel}, \citenamefont {Tamborra}, \citenamefont {Gabici},\ and\
  \citenamefont {Ando}}]{Zandanel:2014pva}%
  \BibitemOpen
  \bibfield  {author} {\bibinfo {author} {\bibfnamefont {F.}~\bibnamefont
  {Zandanel}}, \bibinfo {author} {\bibfnamefont {I.}~\bibnamefont {Tamborra}},
  \bibinfo {author} {\bibfnamefont {S.}~\bibnamefont {Gabici}}, \ and\ \bibinfo
  {author} {\bibfnamefont {S.}~\bibnamefont {Ando}},\ }\href {\doibase
  10.1051/0004-6361/201425249} {\bibfield  {journal} {\bibinfo  {journal}
  {Astron. Astrophys.}\ }\textbf {\bibinfo {volume} {578}},\ \bibinfo {pages}
  {A32} (\bibinfo {year} {2015})},\ \Eprint {http://arxiv.org/abs/1410.8697}
  {arXiv:1410.8697 [astro-ph.HE]} \BibitemShut {NoStop}%
\bibitem [{\citenamefont {Feldstein}\ \emph {et~al.}(2013)\citenamefont
  {Feldstein}, \citenamefont {Kusenko}, \citenamefont {Matsumoto},\ and\
  \citenamefont {Yanagida}}]{Feldstein:2013kka}%
  \BibitemOpen
  \bibfield  {author} {\bibinfo {author} {\bibfnamefont {B.}~\bibnamefont
  {Feldstein}}, \bibinfo {author} {\bibfnamefont {A.}~\bibnamefont {Kusenko}},
  \bibinfo {author} {\bibfnamefont {S.}~\bibnamefont {Matsumoto}}, \ and\
  \bibinfo {author} {\bibfnamefont {T.~T.}\ \bibnamefont {Yanagida}},\ }\href
  {\doibase 10.1103/PhysRevD.88.015004} {\bibfield  {journal} {\bibinfo
  {journal} {Phys. Rev. D}\ }\textbf {\bibinfo {volume} {88}},\ \bibinfo
  {pages} {015004} (\bibinfo {year} {2013})},\ \Eprint
  {http://arxiv.org/abs/1303.7320} {arXiv:1303.7320 [hep-ph]} \BibitemShut
  {NoStop}%
\bibitem [{\citenamefont {Esmaili}\ and\ \citenamefont
  {Serpico}(2013)}]{Esmaili:2013gha}%
  \BibitemOpen
  \bibfield  {author} {\bibinfo {author} {\bibfnamefont {A.}~\bibnamefont
  {Esmaili}}\ and\ \bibinfo {author} {\bibfnamefont {P.~D.}\ \bibnamefont
  {Serpico}},\ }\href {\doibase 10.1088/1475-7516/2013/11/054} {\bibfield
  {journal} {\bibinfo  {journal} {JCAP}\ }\textbf {\bibinfo {volume} {1311}},\
  \bibinfo {pages} {054} (\bibinfo {year} {2013})},\ \Eprint
  {http://arxiv.org/abs/1308.1105} {arXiv:1308.1105 [hep-ph]} \BibitemShut
  {NoStop}%
\bibitem [{\citenamefont {Esmaili}\ \emph {et~al.}(2014)\citenamefont
  {Esmaili}, \citenamefont {Kang},\ and\ \citenamefont
  {Serpico}}]{Esmaili:2014rma}%
  \BibitemOpen
  \bibfield  {author} {\bibinfo {author} {\bibfnamefont {A.}~\bibnamefont
  {Esmaili}}, \bibinfo {author} {\bibfnamefont {S.~K.}\ \bibnamefont {Kang}}, \
  and\ \bibinfo {author} {\bibfnamefont {P.~D.}\ \bibnamefont {Serpico}},\
  }\href {\doibase 10.1088/1475-7516/2014/12/054} {\bibfield  {journal}
  {\bibinfo  {journal} {JCAP}\ }\textbf {\bibinfo {volume} {1412}},\ \bibinfo
  {pages} {054} (\bibinfo {year} {2014})},\ \Eprint
  {http://arxiv.org/abs/1410.5979} {arXiv:1410.5979 [hep-ph]} \BibitemShut
  {NoStop}%
\bibitem [{\citenamefont {Zavala}(2014)}]{Zavala:2014dla}%
  \BibitemOpen
  \bibfield  {author} {\bibinfo {author} {\bibfnamefont {J.}~\bibnamefont
  {Zavala}},\ }\href {\doibase 10.1103/PhysRevD.89.123516} {\bibfield
  {journal} {\bibinfo  {journal} {Phys. Rev. D}\ }\textbf {\bibinfo {volume}
  {89}},\ \bibinfo {pages} {123516} (\bibinfo {year} {2014})},\ \Eprint
  {http://arxiv.org/abs/1404.2932} {arXiv:1404.2932 [astro-ph.HE]} \BibitemShut
  {NoStop}%
\bibitem [{\citenamefont {Murase}\ \emph
  {et~al.}(2015{\natexlab{a}})\citenamefont {Murase}, \citenamefont {Laha},
  \citenamefont {Ando},\ and\ \citenamefont {Ahlers}}]{Murase:2015gea}%
  \BibitemOpen
  \bibfield  {author} {\bibinfo {author} {\bibfnamefont {K.}~\bibnamefont
  {Murase}}, \bibinfo {author} {\bibfnamefont {R.}~\bibnamefont {Laha}},
  \bibinfo {author} {\bibfnamefont {S.}~\bibnamefont {Ando}}, \ and\ \bibinfo
  {author} {\bibfnamefont {M.}~\bibnamefont {Ahlers}},\ }\href {\doibase
  10.1103/PhysRevLett.115.071301} {\bibfield  {journal} {\bibinfo  {journal}
  {Phys. Rev. Lett.}\ }\textbf {\bibinfo {volume} {115}},\ \bibinfo {pages}
  {071301} (\bibinfo {year} {2015}{\natexlab{a}})},\ \Eprint
  {http://arxiv.org/abs/1503.04663} {arXiv:1503.04663 [hep-ph]} \BibitemShut
  {NoStop}%
\bibitem [{\citenamefont {Winter}(2013)}]{Winter:2013cla}%
  \BibitemOpen
  \bibfield  {author} {\bibinfo {author} {\bibfnamefont {W.}~\bibnamefont
  {Winter}},\ }\href {\doibase 10.1103/PhysRevD.88.083007} {\bibfield
  {journal} {\bibinfo  {journal} {Phys. Rev. D}\ }\textbf {\bibinfo {volume}
  {88}},\ \bibinfo {pages} {083007} (\bibinfo {year} {2013})},\ \Eprint
  {http://arxiv.org/abs/1307.2793} {arXiv:1307.2793 [astro-ph.HE]} \BibitemShut
  {NoStop}%
\bibitem [{\citenamefont {Murase}\ \emph {et~al.}(2013)\citenamefont {Murase},
  \citenamefont {Ahlers},\ and\ \citenamefont {Lacki}}]{Murase:2013rfa}%
  \BibitemOpen
  \bibfield  {author} {\bibinfo {author} {\bibfnamefont {K.}~\bibnamefont
  {Murase}}, \bibinfo {author} {\bibfnamefont {M.}~\bibnamefont {Ahlers}}, \
  and\ \bibinfo {author} {\bibfnamefont {B.~C.}\ \bibnamefont {Lacki}},\ }\href
  {\doibase 10.1103/PhysRevD.88.121301} {\bibfield  {journal} {\bibinfo
  {journal} {Phys. Rev. D}\ }\textbf {\bibinfo {volume} {88}},\ \bibinfo
  {pages} {121301} (\bibinfo {year} {2013})},\ \Eprint
  {http://arxiv.org/abs/1306.3417} {arXiv:1306.3417 [astro-ph.HE]} \BibitemShut
  {NoStop}%
\bibitem [{\citenamefont {Ackermann}\ \emph {et~al.}(2015)\citenamefont
  {Ackermann} \emph {et~al.}}]{Ackermann:2014usa}%
  \BibitemOpen
  \bibfield  {author} {\bibinfo {author} {\bibfnamefont {M.}~\bibnamefont
  {Ackermann}} \emph {et~al.} (\bibinfo {collaboration} {Fermi-LAT}),\ }\href
  {\doibase 10.1088/0004-637X/799/1/86} {\bibfield  {journal} {\bibinfo
  {journal} {Astrophys. J.}\ }\textbf {\bibinfo {volume} {799}},\ \bibinfo
  {pages} {86} (\bibinfo {year} {2015})},\ \Eprint
  {http://arxiv.org/abs/1410.3696} {arXiv:1410.3696 [astro-ph.HE]} \BibitemShut
  {NoStop}%
\bibitem [{\citenamefont {Anchordoqui}\ \emph
  {et~al.}(2014{\natexlab{c}})\citenamefont {Anchordoqui}, \citenamefont
  {Goldberg}, \citenamefont {Lynch}, \citenamefont {Olinto}, \citenamefont
  {Paul},\ and\ \citenamefont {Weiler}}]{Anchordoqui:2013qsi}%
  \BibitemOpen
  \bibfield  {author} {\bibinfo {author} {\bibfnamefont {L.~A.}\ \bibnamefont
  {Anchordoqui}}, \bibinfo {author} {\bibfnamefont {H.}~\bibnamefont
  {Goldberg}}, \bibinfo {author} {\bibfnamefont {M.~H.}\ \bibnamefont {Lynch}},
  \bibinfo {author} {\bibfnamefont {A.~V.}\ \bibnamefont {Olinto}}, \bibinfo
  {author} {\bibfnamefont {T.~C.}\ \bibnamefont {Paul}}, \ and\ \bibinfo
  {author} {\bibfnamefont {T.~J.}\ \bibnamefont {Weiler}},\ }\href {\doibase
  10.1103/PhysRevD.89.083003} {\bibfield  {journal} {\bibinfo  {journal} {Phys.
  Rev. D}\ }\textbf {\bibinfo {volume} {89}},\ \bibinfo {pages} {083003}
  (\bibinfo {year} {2014}{\natexlab{c}})},\ \Eprint
  {http://arxiv.org/abs/1306.5021} {arXiv:1306.5021 [astro-ph.HE]} \BibitemShut
  {NoStop}%
\bibitem [{\citenamefont {Fang}\ \emph {et~al.}(2014)\citenamefont {Fang},
  \citenamefont {Fujii}, \citenamefont {Linden},\ and\ \citenamefont
  {Olinto}}]{Fang:2014uja}%
  \BibitemOpen
  \bibfield  {author} {\bibinfo {author} {\bibfnamefont {K.}~\bibnamefont
  {Fang}}, \bibinfo {author} {\bibfnamefont {T.}~\bibnamefont {Fujii}},
  \bibinfo {author} {\bibfnamefont {T.}~\bibnamefont {Linden}}, \ and\ \bibinfo
  {author} {\bibfnamefont {A.~V.}\ \bibnamefont {Olinto}},\ }\href {\doibase
  10.1088/0004-637X/794/2/126} {\bibfield  {journal} {\bibinfo  {journal}
  {Astrophys. J.}\ }\textbf {\bibinfo {volume} {794}},\ \bibinfo {pages} {126}
  (\bibinfo {year} {2014})},\ \Eprint {http://arxiv.org/abs/1404.6237}
  {arXiv:1404.6237 [astro-ph.HE]} \BibitemShut {NoStop}%
\bibitem [{\citenamefont {Kistler}\ \emph {et~al.}(2014)\citenamefont
  {Kistler}, \citenamefont {Stanev},\ and\ \citenamefont
  {Y{\"u}ksel}}]{Kistler:2013my}%
  \BibitemOpen
  \bibfield  {author} {\bibinfo {author} {\bibfnamefont {M.~D.}\ \bibnamefont
  {Kistler}}, \bibinfo {author} {\bibfnamefont {T.}~\bibnamefont {Stanev}}, \
  and\ \bibinfo {author} {\bibfnamefont {H.}~\bibnamefont {Y{\"u}ksel}},\
  }\href {\doibase 10.1103/PhysRevD.90.123006} {\bibfield  {journal} {\bibinfo
  {journal} {Phys. Rev. D}\ }\textbf {\bibinfo {volume} {90}},\ \bibinfo
  {pages} {123006} (\bibinfo {year} {2014})},\ \Eprint
  {http://arxiv.org/abs/1301.1703} {arXiv:1301.1703 [astro-ph.HE]} \BibitemShut
  {NoStop}%
\bibitem [{\citenamefont {Xia}\ \emph {et~al.}(2015)\citenamefont {Xia},
  \citenamefont {Cuoco}, \citenamefont {Branchini},\ and\ \citenamefont
  {Viel}}]{Xia:2015wka}%
  \BibitemOpen
  \bibfield  {author} {\bibinfo {author} {\bibfnamefont {J.-Q.}\ \bibnamefont
  {Xia}}, \bibinfo {author} {\bibfnamefont {A.}~\bibnamefont {Cuoco}}, \bibinfo
  {author} {\bibfnamefont {E.}~\bibnamefont {Branchini}}, \ and\ \bibinfo
  {author} {\bibfnamefont {M.}~\bibnamefont {Viel}},\ }\href {\doibase
  10.1088/0067-0049/217/1/15} {\bibfield  {journal} {\bibinfo  {journal}
  {Astrophys. J. Suppl.}\ }\textbf {\bibinfo {volume} {217}},\ \bibinfo {pages}
  {15} (\bibinfo {year} {2015})},\ \Eprint {http://arxiv.org/abs/1503.05918}
  {arXiv:1503.05918 [astro-ph.CO]} \BibitemShut {NoStop}%
\bibitem [{\citenamefont {Ando}\ \emph {et~al.}(2014)\citenamefont {Ando},
  \citenamefont {Benoit-L{\'e}vy},\ and\ \citenamefont
  {Komatsu}}]{Ando:2013xwa}%
  \BibitemOpen
  \bibfield  {author} {\bibinfo {author} {\bibfnamefont {S.}~\bibnamefont
  {Ando}}, \bibinfo {author} {\bibfnamefont {A.}~\bibnamefont
  {Benoit-L{\'e}vy}}, \ and\ \bibinfo {author} {\bibfnamefont {E.}~\bibnamefont
  {Komatsu}},\ }\href {\doibase 10.1103/PhysRevD.90.023514} {\bibfield
  {journal} {\bibinfo  {journal} {Phys. Rev. D}\ }\textbf {\bibinfo {volume}
  {90}},\ \bibinfo {pages} {023514} (\bibinfo {year} {2014})},\ \Eprint
  {http://arxiv.org/abs/1312.4403} {arXiv:1312.4403 [astro-ph.CO]} \BibitemShut
  {NoStop}%
\bibitem [{\citenamefont {Ando}(2014)}]{Ando:2014aoa}%
  \BibitemOpen
  \bibfield  {author} {\bibinfo {author} {\bibfnamefont {S.}~\bibnamefont
  {Ando}},\ }\href {\doibase 10.1088/1475-7516/2014/10/061} {\bibfield
  {journal} {\bibinfo  {journal} {JCAP}\ }\textbf {\bibinfo {volume} {1410}},\
  \bibinfo {pages} {061} (\bibinfo {year} {2014})},\ \Eprint
  {http://arxiv.org/abs/1407.8502} {arXiv:1407.8502 [astro-ph.CO]} \BibitemShut
  {NoStop}%
\bibitem [{\citenamefont {Fornengo}\ and\ \citenamefont
  {Regis}(2014)}]{Fornengo:2013rga}%
  \BibitemOpen
  \bibfield  {author} {\bibinfo {author} {\bibfnamefont {N.}~\bibnamefont
  {Fornengo}}\ and\ \bibinfo {author} {\bibfnamefont {M.}~\bibnamefont
  {Regis}},\ }\href {\doibase 10.3389/fphy.2014.00006} {\bibfield  {journal}
  {\bibinfo  {journal} {Front. Physics}\ }\textbf {\bibinfo {volume} {2}},\
  \bibinfo {pages} {6} (\bibinfo {year} {2014})},\ \Eprint
  {http://arxiv.org/abs/1312.4835} {arXiv:1312.4835 [astro-ph.CO]} \BibitemShut
  {NoStop}%
\bibitem [{\citenamefont {Cuoco}\ \emph {et~al.}(2015)\citenamefont {Cuoco},
  \citenamefont {Xia}, \citenamefont {Regis}, \citenamefont {Branchini},
  \citenamefont {Fornengo},\ and\ \citenamefont {Viel}}]{Cuoco:2015rfa}%
  \BibitemOpen
  \bibfield  {author} {\bibinfo {author} {\bibfnamefont {A.}~\bibnamefont
  {Cuoco}}, \bibinfo {author} {\bibfnamefont {J.-Q.}\ \bibnamefont {Xia}},
  \bibinfo {author} {\bibfnamefont {M.}~\bibnamefont {Regis}}, \bibinfo
  {author} {\bibfnamefont {E.}~\bibnamefont {Branchini}}, \bibinfo {author}
  {\bibfnamefont {N.}~\bibnamefont {Fornengo}}, \ and\ \bibinfo {author}
  {\bibfnamefont {M.}~\bibnamefont {Viel}},\ }\href@noop {} {\  (\bibinfo
  {year} {2015})},\ \Eprint {http://arxiv.org/abs/1506.01030} {arXiv:1506.01030
  [astro-ph.HE]} \BibitemShut {NoStop}%
\bibitem [{\citenamefont {Ade}\ \emph {et~al.}(2015)\citenamefont {Ade} \emph
  {et~al.}}]{Ade:2015xua}%
  \BibitemOpen
  \bibfield  {author} {\bibinfo {author} {\bibfnamefont {P.~A.~R.}\
  \bibnamefont {Ade}} \emph {et~al.} (\bibinfo {collaboration} {Planck}),\
  }\href@noop {} {\  (\bibinfo {year} {2015})},\ \Eprint
  {http://arxiv.org/abs/1502.01589} {arXiv:1502.01589 [astro-ph.CO]}
  \BibitemShut {NoStop}%
\bibitem [{\citenamefont {Gruppioni}\ \emph {et~al.}(2013)\citenamefont
  {Gruppioni} \emph {et~al.}}]{Gruppioni:2013jna}%
  \BibitemOpen
  \bibfield  {author} {\bibinfo {author} {\bibfnamefont {C.}~\bibnamefont
  {Gruppioni}} \emph {et~al.},\ }\href {\doibase 10.1093/mnras/stt308}
  {\bibfield  {journal} {\bibinfo  {journal} {Mon. Not. R. Astron. Soc.}\
  }\textbf {\bibinfo {volume} {432}},\ \bibinfo {pages} {23} (\bibinfo {year}
  {2013})},\ \Eprint {http://arxiv.org/abs/1302.5209} {arXiv:1302.5209
  [astro-ph.CO]} \BibitemShut {NoStop}%
\bibitem [{\citenamefont {Finke}\ \emph {et~al.}(2010)\citenamefont {Finke},
  \citenamefont {Razzaque},\ and\ \citenamefont {Dermer}}]{Finke:2009xi}%
  \BibitemOpen
  \bibfield  {author} {\bibinfo {author} {\bibfnamefont {J.~D.}\ \bibnamefont
  {Finke}}, \bibinfo {author} {\bibfnamefont {S.}~\bibnamefont {Razzaque}}, \
  and\ \bibinfo {author} {\bibfnamefont {C.~D.}\ \bibnamefont {Dermer}},\
  }\href {\doibase 10.1088/0004-637X/712/1/238} {\bibfield  {journal} {\bibinfo
   {journal} {Astrophys. J.}\ }\textbf {\bibinfo {volume} {712}},\ \bibinfo
  {pages} {238} (\bibinfo {year} {2010})},\ \Eprint
  {http://arxiv.org/abs/0905.1115} {arXiv:0905.1115 [astro-ph.HE]} \BibitemShut
  {NoStop}%
\bibitem [{\citenamefont {Blas}\ \emph {et~al.}(2011)\citenamefont {Blas},
  \citenamefont {Lesgourgues},\ and\ \citenamefont {Tram}}]{Blas:2011rf}%
  \BibitemOpen
  \bibfield  {author} {\bibinfo {author} {\bibfnamefont {D.}~\bibnamefont
  {Blas}}, \bibinfo {author} {\bibfnamefont {J.}~\bibnamefont {Lesgourgues}}, \
  and\ \bibinfo {author} {\bibfnamefont {T.}~\bibnamefont {Tram}},\ }\href
  {\doibase 10.1088/1475-7516/2011/07/034} {\bibfield  {journal} {\bibinfo
  {journal} {JCAP}\ }\textbf {\bibinfo {volume} {1107}},\ \bibinfo {pages}
  {034} (\bibinfo {year} {2011})},\ \Eprint {http://arxiv.org/abs/1104.2933}
  {arXiv:1104.2933 [astro-ph.CO]} \BibitemShut {NoStop}%
\bibitem [{\citenamefont {Ando}\ and\ \citenamefont
  {Kusenko}(2010)}]{Ando:2010rb}%
  \BibitemOpen
  \bibfield  {author} {\bibinfo {author} {\bibfnamefont {S.}~\bibnamefont
  {Ando}}\ and\ \bibinfo {author} {\bibfnamefont {A.}~\bibnamefont {Kusenko}},\
  }\href {\doibase 10.1088/2041-8205/722/1/L39} {\bibfield  {journal} {\bibinfo
   {journal} {Astrophys. J.}\ }\textbf {\bibinfo {volume} {722}},\ \bibinfo
  {pages} {L39} (\bibinfo {year} {2010})},\ \Eprint
  {http://arxiv.org/abs/1005.1924} {arXiv:1005.1924 [astro-ph.HE]} \BibitemShut
  {NoStop}%
\bibitem [{\citenamefont {Anchordoqui}\ \emph {et~al.}(2004)\citenamefont
  {Anchordoqui}, \citenamefont {Goldberg}, \citenamefont {Halzen},\ and\
  \citenamefont {Weiler}}]{Anchordoqui:2004eu}%
  \BibitemOpen
  \bibfield  {author} {\bibinfo {author} {\bibfnamefont {L.~A.}\ \bibnamefont
  {Anchordoqui}}, \bibinfo {author} {\bibfnamefont {H.}~\bibnamefont
  {Goldberg}}, \bibinfo {author} {\bibfnamefont {F.}~\bibnamefont {Halzen}}, \
  and\ \bibinfo {author} {\bibfnamefont {T.~J.}\ \bibnamefont {Weiler}},\
  }\href {\doibase 10.1016/j.physletb.2004.09.005} {\bibfield  {journal}
  {\bibinfo  {journal} {Phys. Lett.}\ }\textbf {\bibinfo {volume} {B600}},\
  \bibinfo {pages} {202} (\bibinfo {year} {2004})},\ \Eprint
  {http://arxiv.org/abs/astro-ph/0404387} {arXiv:astro-ph/0404387 [astro-ph]}
  \BibitemShut {NoStop}%
\bibitem [{\citenamefont {Ackermann}\ \emph {et~al.}(2012)\citenamefont
  {Ackermann} \emph {et~al.}}]{Ackermann:2012vca}%
  \BibitemOpen
  \bibfield  {author} {\bibinfo {author} {\bibfnamefont {M.}~\bibnamefont
  {Ackermann}} \emph {et~al.} (\bibinfo {collaboration} {Fermi-LAT}),\ }\href
  {\doibase 10.1088/0004-637X/755/2/164} {\bibfield  {journal} {\bibinfo
  {journal} {Astrophys. J.}\ }\textbf {\bibinfo {volume} {755}},\ \bibinfo
  {pages} {164} (\bibinfo {year} {2012})},\ \Eprint
  {http://arxiv.org/abs/1206.1346} {arXiv:1206.1346 [astro-ph.HE]} \BibitemShut
  {NoStop}%
\bibitem [{\citenamefont {Ajello}\ \emph {et~al.}(2015)\citenamefont {Ajello}
  \emph {et~al.}}]{Ajello:2015mfa}%
  \BibitemOpen
  \bibfield  {author} {\bibinfo {author} {\bibfnamefont {M.}~\bibnamefont
  {Ajello}} \emph {et~al.},\ }\href {\doibase 10.1088/2041-8205/800/2/L27}
  {\bibfield  {journal} {\bibinfo  {journal} {Astrophys. J.}\ }\textbf
  {\bibinfo {volume} {800}},\ \bibinfo {pages} {L27} (\bibinfo {year}
  {2015})},\ \Eprint {http://arxiv.org/abs/1501.05301} {arXiv:1501.05301
  [astro-ph.HE]} \BibitemShut {NoStop}%
\bibitem [{\citenamefont {Tinker}\ \emph {et~al.}(2010)\citenamefont {Tinker},
  \citenamefont {Robertson}, \citenamefont {Kravtsov}, \citenamefont {Klypin},
  \citenamefont {Warren}, \citenamefont {Yepes},\ and\ \citenamefont
  {Gottlober}}]{Tinker:2010my}%
  \BibitemOpen
  \bibfield  {author} {\bibinfo {author} {\bibfnamefont {J.~L.}\ \bibnamefont
  {Tinker}}, \bibinfo {author} {\bibfnamefont {B.~E.}\ \bibnamefont
  {Robertson}}, \bibinfo {author} {\bibfnamefont {A.~V.}\ \bibnamefont
  {Kravtsov}}, \bibinfo {author} {\bibfnamefont {A.}~\bibnamefont {Klypin}},
  \bibinfo {author} {\bibfnamefont {M.~S.}\ \bibnamefont {Warren}}, \bibinfo
  {author} {\bibfnamefont {G.}~\bibnamefont {Yepes}}, \ and\ \bibinfo {author}
  {\bibfnamefont {S.}~\bibnamefont {Gottlober}},\ }\href {\doibase
  10.1088/0004-637X/724/2/878} {\bibfield  {journal} {\bibinfo  {journal}
  {Astrophys. J.}\ }\textbf {\bibinfo {volume} {724}},\ \bibinfo {pages} {878}
  (\bibinfo {year} {2010})},\ \Eprint {http://arxiv.org/abs/1001.3162}
  {arXiv:1001.3162 [astro-ph.CO]} \BibitemShut {NoStop}%
\bibitem [{\citenamefont {Murase}\ \emph
  {et~al.}(2015{\natexlab{b}})\citenamefont {Murase}, \citenamefont {Guetta},\
  and\ \citenamefont {Ahlers}}]{Murase:2015xka}%
  \BibitemOpen
  \bibfield  {author} {\bibinfo {author} {\bibfnamefont {K.}~\bibnamefont
  {Murase}}, \bibinfo {author} {\bibfnamefont {D.}~\bibnamefont {Guetta}}, \
  and\ \bibinfo {author} {\bibfnamefont {M.}~\bibnamefont {Ahlers}},\
  }\href@noop {} {\  (\bibinfo {year} {2015}{\natexlab{b}})},\ \Eprint
  {http://arxiv.org/abs/1509.00805} {arXiv:1509.00805 [astro-ph.HE]}
  \BibitemShut {NoStop}%
\bibitem [{\citenamefont {Camera}\ \emph {et~al.}(2015)\citenamefont {Camera},
  \citenamefont {Fornasa}, \citenamefont {Fornengo},\ and\ \citenamefont
  {Regis}}]{Camera:2014rja}%
  \BibitemOpen
  \bibfield  {author} {\bibinfo {author} {\bibfnamefont {S.}~\bibnamefont
  {Camera}}, \bibinfo {author} {\bibfnamefont {M.}~\bibnamefont {Fornasa}},
  \bibinfo {author} {\bibfnamefont {N.}~\bibnamefont {Fornengo}}, \ and\
  \bibinfo {author} {\bibfnamefont {M.}~\bibnamefont {Regis}},\ }\href
  {\doibase 10.1088/1475-7516/2015/06/029} {\bibfield  {journal} {\bibinfo
  {journal} {JCAP}\ }\textbf {\bibinfo {volume} {1506}},\ \bibinfo {pages}
  {029} (\bibinfo {year} {2015})},\ \Eprint {http://arxiv.org/abs/1411.4651}
  {arXiv:1411.4651 [astro-ph.CO]} \BibitemShut {NoStop}%
\bibitem [{\citenamefont {Shirasaki}\ \emph {et~al.}(2014)\citenamefont
  {Shirasaki}, \citenamefont {Horiuchi},\ and\ \citenamefont
  {Yoshida}}]{Shirasaki:2014noa}%
  \BibitemOpen
  \bibfield  {author} {\bibinfo {author} {\bibfnamefont {M.}~\bibnamefont
  {Shirasaki}}, \bibinfo {author} {\bibfnamefont {S.}~\bibnamefont {Horiuchi}},
  \ and\ \bibinfo {author} {\bibfnamefont {N.}~\bibnamefont {Yoshida}},\ }\href
  {\doibase 10.1103/PhysRevD.90.063502} {\bibfield  {journal} {\bibinfo
  {journal} {Phys. Rev. D}\ }\textbf {\bibinfo {volume} {90}},\ \bibinfo
  {pages} {063502} (\bibinfo {year} {2014})},\ \Eprint
  {http://arxiv.org/abs/1404.5503} {arXiv:1404.5503 [astro-ph.CO]} \BibitemShut
  {NoStop}%
\bibitem [{\citenamefont {Fornengo}\ \emph {et~al.}(2015)\citenamefont
  {Fornengo}, \citenamefont {Perotto}, \citenamefont {Regis},\ and\
  \citenamefont {Camera}}]{Fornengo:2014cya}%
  \BibitemOpen
  \bibfield  {author} {\bibinfo {author} {\bibfnamefont {N.}~\bibnamefont
  {Fornengo}}, \bibinfo {author} {\bibfnamefont {L.}~\bibnamefont {Perotto}},
  \bibinfo {author} {\bibfnamefont {M.}~\bibnamefont {Regis}}, \ and\ \bibinfo
  {author} {\bibfnamefont {S.}~\bibnamefont {Camera}},\ }\href {\doibase
  10.1088/2041-8205/802/1/L1} {\bibfield  {journal} {\bibinfo  {journal}
  {Astrophys. J.}\ }\textbf {\bibinfo {volume} {802}},\ \bibinfo {pages} {L1}
  (\bibinfo {year} {2015})},\ \Eprint {http://arxiv.org/abs/1410.4997}
  {arXiv:1410.4997 [astro-ph.CO]} \BibitemShut {NoStop}%
\end{thebibliography}%

\end{document}